\documentclass[twocolumn]{aastex63}
\pdfoutput=1

\newcommand\Gaia{{\it Gaia }}

\usepackage{amsmath,graphicx,multirow,booktabs}

\received{}
\revised{}
\accepted{}
\submitjournal{ApJ}

\shorttitle{Photometric Classifications of Evolved Massive Stars}
\shortauthors{Dorn-Wallenstein, Davenport, Huppenkothen, \& Levesque}

\begin{document}

\title{Photometric Classifications of Evolved Massive Stars: \\ Preparing for the Era of {\it Webb} and {\it Roman} with Machine Learning}

\correspondingauthor{Trevor Z. Dorn-Wallenstein}
\email{tzdw@uw.edu}

\author[0000-0003-3601-3180]{Trevor Z. Dorn-Wallenstein}
\affiliation{University of Washington Astronomy Department \\
Physics and Astronomy Building, 3910 15th Ave NE  \\
Seattle, WA 98105, USA} 

\author[0000-0002-0637-835X]{James R. A. Davenport}
\affiliation{University of Washington Astronomy Department \\
Physics and Astronomy Building, 3910 15th Ave NE  \\
Seattle, WA 98105, USA} 

\author[0000-0002-1169-7486]{Daniela Huppenkothen}
\affiliation{SRON Netherlands Institute for Space Research,  \\
Sorbonnelaan 3, 3584 CA Utrecht\\
The Netherlands}

\affiliation{DIRAC Institute,  \\
Department of Astronomy,\\
University of Washington, 3910 15th Ave NE  \\
Seattle, WA 98105, USA} 
\affiliation{The University of Washington eScience Institute,\\
The Washington Research Foundation Data Science Studio, \\
University of Washington, \\
Seattle, WA 98105, USA}

\author[0000-0003-2184-1581]{Emily M. Levesque}
\affiliation{University of Washington Astronomy Department \\
Physics and Astronomy Building, 3910 15th Ave NE  \\
Seattle, WA 98105, USA}

\begin{abstract}

In the coming years, next-generation space-based infrared observatories will significantly increase our samples of rare massive stars, representing a tremendous opportunity to leverage modern statistical tools and methods to test massive stellar evolution in entirely new environments. Such work is only possible if the observed objects can be reliably classified. Spectroscopic observations are infeasible with more distant targets, and so we wish to determine whether machine learning methods can classify massive stars using broadband infrared photometry. We find that a Support Vector Machine classifier is capable of coarsely classifying massive stars with labels corresponding to hot, cool, and emission line stars with high accuracy, while rejecting contaminating low mass giants. Remarkably, 76\% of emission line stars can be recovered without the need for narrowband or spectroscopic observations. We classify a sample of ${\sim}2500$ objects with no existing labels, and identify fourteen candidate emission line objects. Unfortunately, despite the high precision of the photometry in our sample, the heterogeneous origins of the labels for the stars in our sample severely inhibits our classifier from distinguishing classes of stars with more granularity. Ultimately, no large and homogeneously labeled sample of massive stars currently exists. Without significant efforts to robustly classify evolved massive stars --- which is feasible given existing data from large all-sky spectroscopic surveys --- shortcomings in the labeling of existing data sets will hinder efforts to leverage the next-generation of space observatories.

\end{abstract}


\section{Introduction} \label{sec:intro}

Evolved massive stars are observed in a menagerie of exotic evolutionary phases. While the challenge of connecting these states with a self-consistent theory of stellar evolution has seen rapid advancement since the original introduction of the ``Conti Scenario'' \citep{conti83}, the effects of rotation, magnetic fields, internal mixing processes, and binary interactions on the evolution of massive stars are still the subject of much theoretical effort \citep[e.g.,][]{ekstrom12,eldridge17}. While individual massive stars can be used as precision probes of these processes, {\it ensembles} of evolved massive stars can also significantly constrain stellar evolution. This can be done by comparing the integrated spectra of massive stars \citep[e.g.][]{levesque12}, or by studying the detailed makeup of resolved populations of massive stars \citep{dornwallenstein18,dornwallenstein20,stanway20}. 

Using the demographics of stellar populations to constrain stellar evolution requires large and accurately-classified samples of evolved massive stars. Such samples will be achievable in the coming years with the launch of the James Webb Space Telescope ({\it Webb}) and the Nancy Grace Roman Space Telescope ({\it Roman}). Among the instrumentation on {\it Webb} and the proposed instrumentation for {\it Roman} are photometers equipped with filters spanning a broad wavelength baseline from 0.5 to 28 $\mu$m. The resolution of {\it Webb} will allow us to identify and study in detail individual luminous stars to great distances \citep[e.g.][]{jones17}, while the impressive 0.218 deg$^{2}$ field of view of {\it Roman} will allow us to efficiently survey nearby galaxies in a small number of pointings \citep{spergel13}. Combined, observations from both missions will give astronomers access to precise infrared measurements of vast numbers of evolved massive stars. But without sophisticated methods of identifying and classifying these stars, the science return afforded by such a large increase in expected sample sizes will be significantly reduced.

Classification of stars from broadband photometry is often done by adopting simple linear cuts in color-magnitude space \citep[e.g.,][]{massey06,massey09}, and --- most critically --- do not include rare emission line objects, whose classification requires dedicated narrow-band surveys (sometimes with custom-designed filters, e.g. \citealt{neugent18}), often accompanied by follow-up spectroscopy, both of which require extensive telescope time. These objects are often the post-main sequence evolved states of massive stars, in which the effects of rotation, binary interactions, and chemical mixing are the most pronounced; the stars that place the most valuable constraints on unknown stellar physics are also the hardest to detect via traditional means. Therefore, it is worthwhile to determine whether there are alternative ways to classify massive stars that avoid using traditional and expensive methods.

At present, we can mimic the observing capabilities of {\it Webb} and {\it Roman} by combining data from \Gaia (which has a red-optical bandpass), the Two Micron All Sky Survey (2MASS, \citealt{skrutskie06},  near-infrared), and the Wide Field Infrared Survey Explorer (WISE, \citealt{wright10}, mid-infrared). WISE provides the additional benefit of having scanned the sky approximately every six months, yielding lightcurves spanning a $\sim7$-year baseline from which we can extract variability metrics for most stars observed. While {\it Roman} and {\it Webb} will not be observing the entire sky in this fashion, determining whether variability can aid in the classification of evolved massive stars will determine whether observers should seek repeated observations of a stellar population.

We wish to determine whether we can
\begin{enumerate}
    \item Assemble a sample of evolved massive stars with available classifications as a training data set,
    \item Construct a machine learning classifier that can reject low mass red contaminants and identify likely emission line objects in order to optimise available telescope time on the most promising targets,
    \item Determine whether variability metrics estimated from WISE lightcurves can aid in these tasks, and
    \item Determine which photometric bandpasses and variability metrics contribute the most to making accurate classifications.
\end{enumerate}

Here we utilize a support vector machine classifier (SVC) trained only on broad-band photometry and simple metrics derived from WISE lightcurves to classify a large sample of evolved massive stars. We describe our sample selection and labeling method in \S\ref{sec:sample}. \S\ref{sec:variabilitymetrics} details the calculation of the simple metrics derived from the WISE lightcurves, and describes the overall behavior of the stars in our sample. We explain our classification algorithm, discuss its successes and shortcomings in \S\ref{sec:svm}, and apply it to a training sample of 2500 stars before presenting our recommendations and concluding in \S\ref{sec:conclusion}.

\section{Sample Selection \& Labeling}\label{sec:sample}

\begin{figure}[ht!]
\plotone{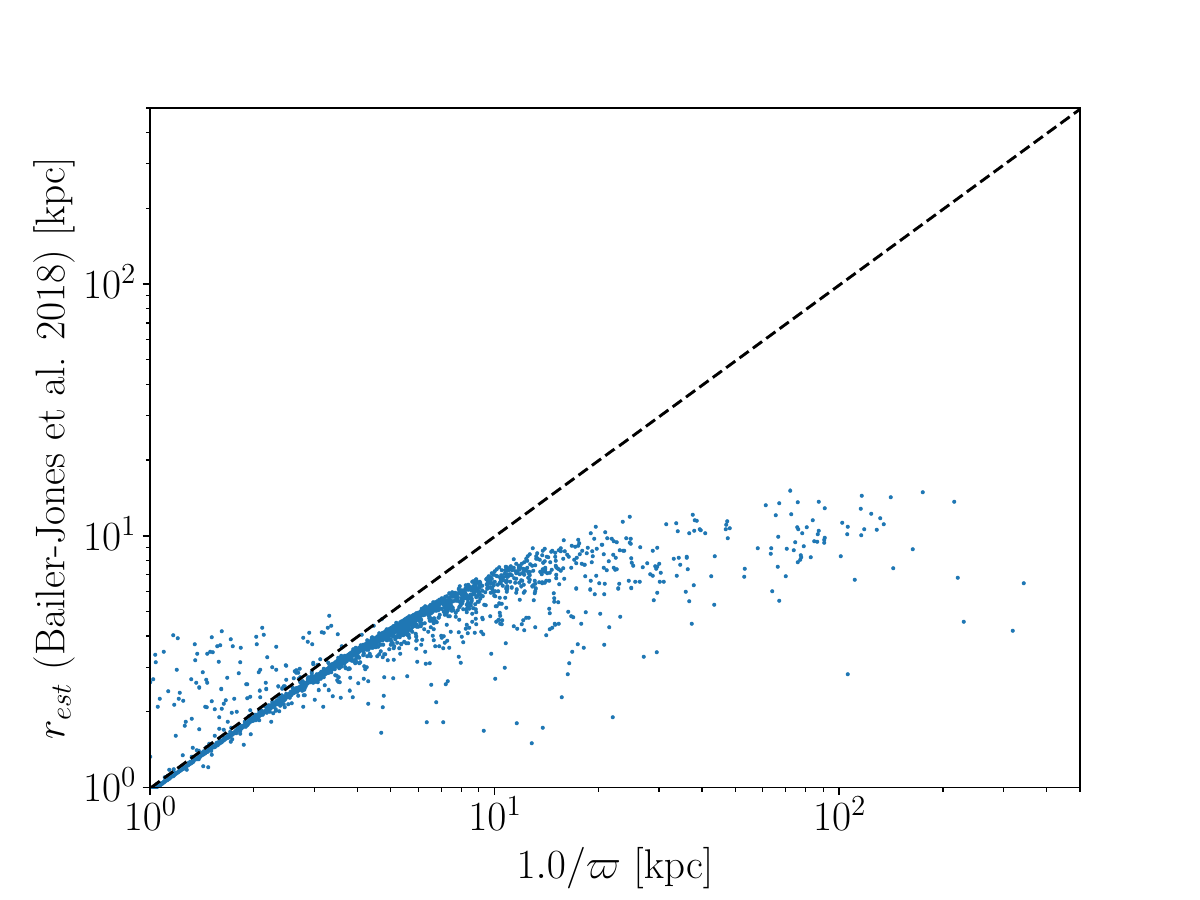}
\caption{Distance from \citet{bailerjones18} versus distance inferred via inverting the reported $\varpi$ from \Gaia DR2 for $\sim10000$ putative massive stars. The dashed line shows where $r_{est}=1/\varpi$.}\label{fig:dist_invparallax}
\end{figure}

For any machine learning algorithm, a high-quality training set with accurate labels is necessary. The second data release (DR2) of the \Gaia mission \citep{gaia18} contains precise photometry in three bands ($G$, $G_{BP}$, and $G_{RP}$) and geometric parallaxes ($\varpi$) for 1.3 billion stars in the Milky Way (MW) and Magellanic Clouds. Because the parallax measurements suffer from some systematics \citep{lindegren18}, and many objects have high fractional errors ($\sigma_\varpi/\varpi$) or negative measured parallax, \citet{bailerjones18} calculated Bayesian distance estimates for the majority of stars in \Gaia DR2, using a prior based on the spatial distribution of stars in the MW. Figure \ref{fig:dist_invparallax} shows the difference between the distance inferred by \citet{bailerjones18}, $r_{est}$, and a naive distance derived by inverting the reported \Gaia measurements of $\varpi$ for $\sim10,000$ putative massive stars (as described below). The dashed line indicates where $r_{est}=1/\varpi$. While the two distance estimates are roughly consistent for nearby stars, more distant stars are biased much further away in the naive distance estimates.

We first perform a cross-match between the \citet{bailerjones18} catalog and the existing cross-match between \Gaia DR2 and the ALLWISE data release. ALLWISE \citep{cutri13} contains photometry in four mid-infrared (MIR) bands --- $W1$ (3.4 $\mu$m), $W2$ (4.6 $\mu$m), $W3$ (12 $\mu$m), and $W4$ (22 $\mu$m) --- derived from co-added images obtained during the original WISE mission, as well as $W1$ and $W2$ images obtained in the post-cryogenic NEOWISE mission \citep{mainzer11}. We select all stars with successful distance estimates (i.e., where ${\tt result\_flag}=1$ if the distance estimate is the mode of the posterior distribution, 2 if it is the median, and 0 for a failed estimate, see \citealt{bailerjones18} for more details) that satisfy

\begin{align}
\begin{split}
M_G &= G - 5\log{r_{est}} + 5 \leq -1.5, \\ 
W1 &< 14.
\end{split}
\end{align}
Since \citet{bailerjones18} used a Galactic prior, stars in the Large and Small Magellanic Clouds (LMC/SMC respectively) have distances that are considerably underestimated. Thus, we also match the catalog in \citet{gaiacollab18} to the ALLWISE/\Gaia cross-match, and select stars with $W1 < 14$ and $M_G \leq -1.5$, assuming distance moduli of 19.05/18.52 for the SMC/LMC respectively \citep{kovacs00a,kovacs00} and combining the two cross-matches while dropping duplicate stars. This results in a total of 452,283 stars.

We then estimate the reddening in the \Gaia bandpasses using the published estimate for $A_G$ from \Gaia DR2, and coefficients from \citep{malhan18} to calculate $E(G_{BP}-G_{RP})$. For Galactic stars without $A_G$ estimates, we assume $A_G = 0$, and for stars in the Magellanic Clouds, we assume the average value of $A_G$ and $E(G_{BP}-G_{RP})$ using $R_V$ measurements from \citet{gordon03} and $E(B-V)$ from \citet{massey07}. Using these quanitites, we calculate the intrinsic $G_{BP}-G_{RP}$ and $M_G$ for all stars.

\begin{figure*}[ht!]
\plotone{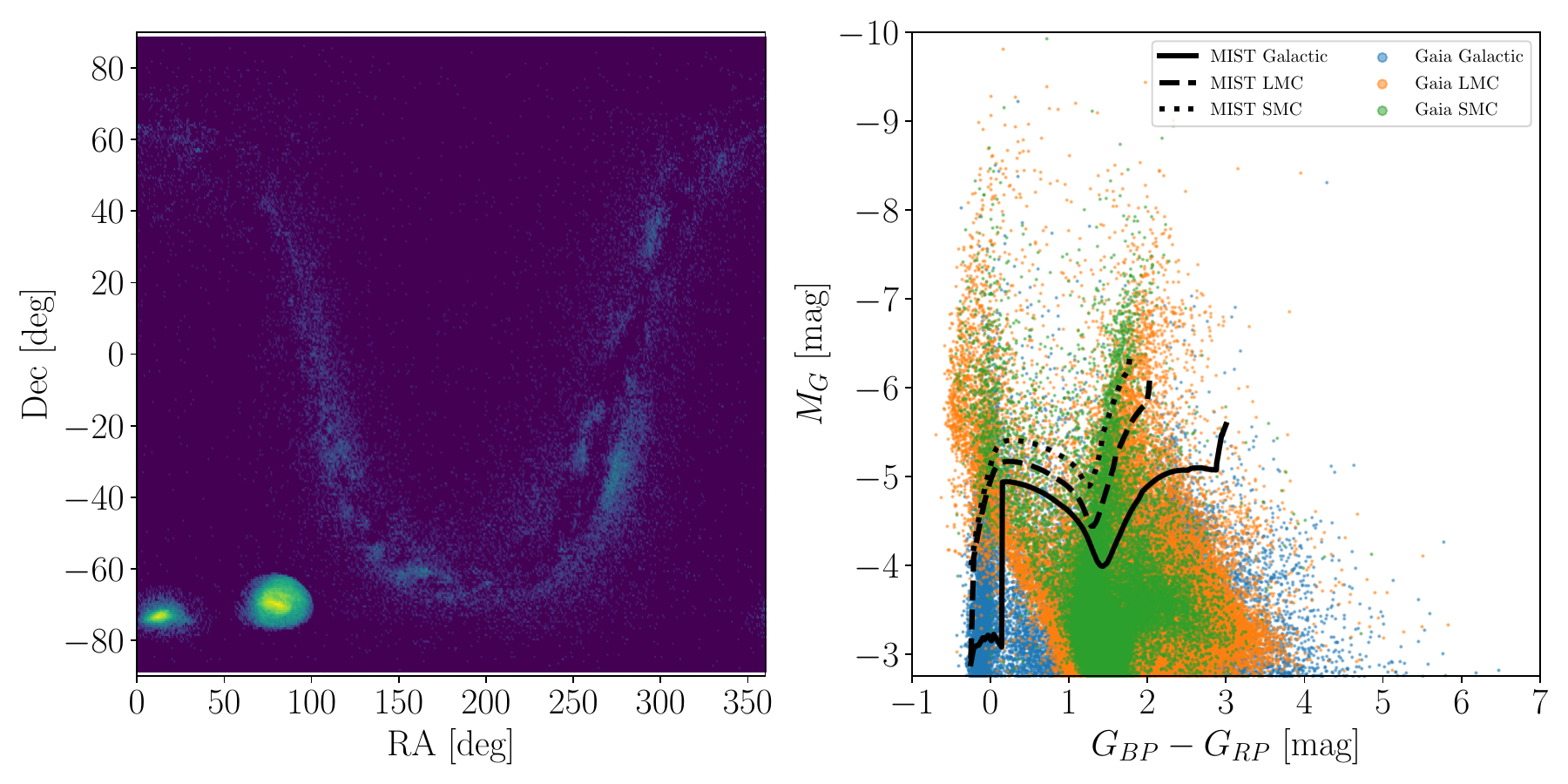}
\caption{{\it Left:} Density of stars selected from the \Gaia database on the sky. Intensity of the colormap corresponds to the logarithm of the number of stars in each bin. {\it Right}: \Gaia CMD for stars selected from the \Gaia DR2 database brighter than $M_G = -1.5$ (though we only plot stars brighter than $M_G = -2.75$ to highlight the likely massive stars). Galactic stars are in blue, LMC stars are in orange, and SMC stars are in green. The solid, dashed, and dotted lines represent our minimum-luminosity criteria to select massive stars in the Galaxy, LMC, and SMC respectively.}\label{fig:skycmd}
\end{figure*}

We can then construct color-magnitude diagrams (CMDs) in the \Gaia filters, which we can use to select massive stars --- i.e., stars with initial mass $M_i \geq 8 M_\odot$. We use the MESA Isochrones \& Stellar Tracks (MIST, \citealt{dotter16,choi16,paxton11,paxton13,paxton15}) isochrones with metallicity $[Fe/H] = 0,-0.5, -1$ for the Galaxy, LMC, and SMC respectively, and rotation speed relative to critical of $v/v_{crit}=0.4$. We then selected the faintest isochrone point of any age with $M_i \geq 8$ M$_\odot$ in 100 equally-spaced bins in the range $-0.25 \leq G_{BP}-G_{RP} \leq 3$. We note that the oldest MIST time bin is $10^{10.3}$ yr (older than the age of the Universe), but by selecting points with $M_i \geq 8$ M$_\odot$, none of the selected points are older than $\sim$40 Myr. These isochrone points form a boundary in the \Gaia CMD that represents the faintest luminosities reached by any massive star at any point during its evolution, and no fainter massive stars are expected to be found --- note that many isochrone points with $M_i < 8$ M$_\odot$ lie above this boundary, so our sample is not constructed to be free of contamination. The left panel of Figure \ref{fig:skycmd} shows the logarithmic density on the sky of all stars selected from the \Gaia DR2 database; the Galactic plane and Magellanic Clouds are clearly visible. The right panel shows the \Gaia CMD, zoomed in show only stars brighter than $M_G=-2.75$, where blue (orange, green) points are individual stars in the Galaxy (LMC, SMC). The solid (dashed, dotted) black line show the MIST luminosity threshold for the Galaxy (LMC, SMC). Note that the thresholds accurately capture the slope of the main sequence for all three galaxies, as well as the $G_{BP}-G_{RP}$ color corresponding to the Hayashi limit.

We select all stars brighter than the corresponding luminosity threshold for their host galaxy, resulting in 9784 objects. From this sample, we select all stars fainter than the saturation limit in $W1$ (8) and $W2$ (7) with valid measurements listed in the ALLWISE catalog for the three bluest WISE bands (excluding $W4$, where the signal-to-noise is often poor). We also convert the $W1$ and $W2$ magnitudes (and uncertainties) to fluxes, and filter for stars with signal-to-noise ratio greater than 3. This results in a final sample of 6484 objects. Table \ref{tab:sample} lists the names, coordinates, host galaxies, distances from \citet{bailerjones18}, and \Gaia photometry for these stars. We query ${\tt Vizier}$ \citep{ochsenbein00} using ${\tt astroquery}$ to download $JHK_s$ photometry from the 2-micron All Sky Survey (2MASS, \citealt{skrutskie06}) for all stars. We also query SIMBAD \citep{wenger00} and download the common name (${\tt MAIN\_ID}$), spectral type (contained in the ${\tt MK\_Spectral\_Type}$ and ${\tt SP\_Type}$ fields), and object type (${\tt OType}$) for each star, the latter two of which we use to assign labels.

\begin{deluxetable*}{lccccccc}
\tabletypesize{\scriptsize}
\tablecaption{Common names, coordinates, host galaxies, and \Gaia measurements of 6,484 putative massive stars, ordered by Right Ascension. $r_{est}$ from \citet{bailerjones18} is given for Galactic stars. Listed values of $G$ and $G_{BP}-G_{RP}$ are uncorrected for extinction.\label{tab:sample}}
\tablehead{\colhead{Common Name} & \colhead{R.A. [deg]} & \colhead{Dec [deg]} & \colhead{Host Galaxy} & \colhead{$r_{est}$ [kpc]} & \colhead{$G$ [mag]} & \colhead{$A_G$ [mag]} & \colhead{$G_{BP}-G_{RP}$ [mag]}} 
\startdata
HD 236270 & 0.17442287 & $55.72245665$ & MW & 2.162 & $9.07$ & $0.94$ & $0.24$ \\ 
LS   I +64   10 & 0.38838658 & $64.51219232$ & MW & 5.305 & $11.55$ & $1.33$ & $0.57$ \\ 
LS   I +60   69 & 0.55506135 & $60.43828347$ & MW & 5.866 & $11.85$ & $1.23$ & $0.61$ \\ 
BD+62  2353 & 0.59458742 & $62.90087875$ & MW & 5.243 & $9.81$ & $0.48$ & $0.37$ \\ 
HD     73 & 1.40408512 & $43.40139506$ & MW & 1.869 & $8.19$ & $0.12$ & $-0.15$ \\ 
HD 240496 & 1.42175475 & $58.49541068$ & MW & 2.499 & $9.70$ & $1.55$ & $0.68$ \\ 
WISE J000559.28-790653.3 & 1.49713706 & $-79.11483482$ & SMC &  -  & $13.94$ & $0.21$ & $1.05$ \\ 
LS   I +59   30 & 1.70503555 & $59.85955733$ & MW & 4.006 & $10.86$ & $1.19$ & $0.50$ \\ 
BD+57  2870 & 1.82960982 & $58.33785301$ & MW & 3.893 & $9.84$ & $1.42$ & $0.82$ \\ 
BD+62     1 & 1.88805102 & $63.08030731$ & MW & 2.893 & $10.29$ & $1.30$ & $0.53$ \\ 
\enddata
\tablecomments{This table is published in its entirety in the machine-readable format. A portion is shown here for guidance regarding its form and content.}
\end{deluxetable*}

\subsection{Label Assignment}

For our final sample of $\sim6500$ stars, we wish to assign the best available estimate of its evolutionary state. These labels can be used to compare to the predictions of stellar population models. Note that these evolutionary states (which are theoretical concepts) are mostly tied to spectral appearance (which is an observable quantity). Therefore we are assuming that, e.g., all stars with Wolf-Rayet spectra are in the same evolutionary state (namely, the descendants of massive stars with high luminosities that have lost their envelopes via strong winds), and that all stars in that evolutionary state are observed as Wolf-Rayet stars. We know that at least the former isn't true, as some stars lose their envelopes due to interactions with a binary companion \citep{eldridge17}, and the latter is also questionable since such stars may or may not appear similar to classical Wolf-Rayets \citep{gotberg18}. Nevertheless, we assume that the assigned labels are a reasonable approximation for a star's evolutionary state with this caveat in mind.

\begin{figure*}[ht!]
\includegraphics[scale=0.55,angle=90]{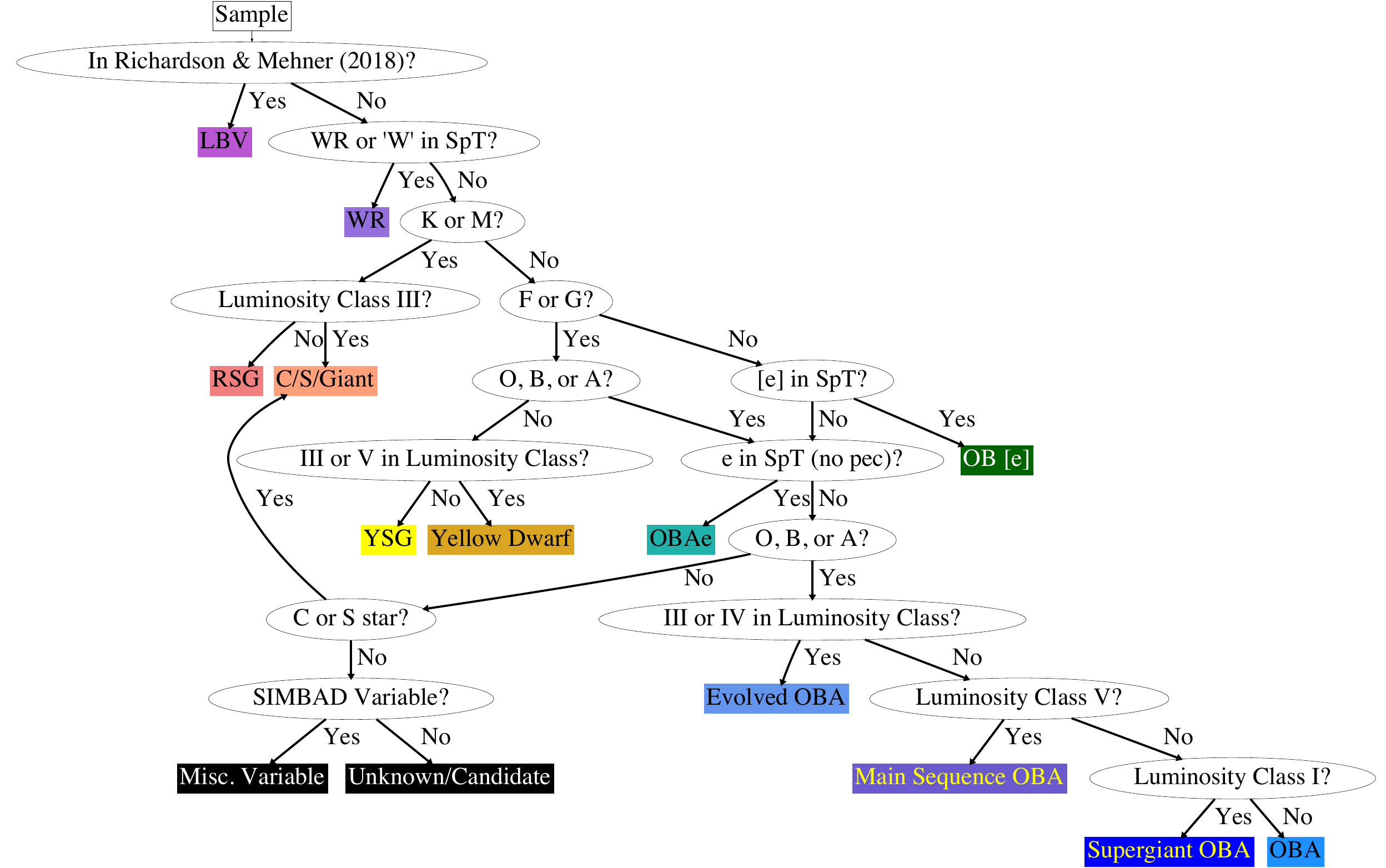}
\caption{Flowchart illustrating the process by which stars are assigned labels, as described in text. Each star begins in the top left and is assigned a label by following a series of binary decisions. This process is complex, and demonstrates the difficulty in deriving useful labels for massive stars. For example, some stars with F or G in their spectral types are actually hot OBA stars (as described in text), and require special handling.}\label{fig:classify}
\end{figure*}

At present, a database of homogeneously classified massive stars does not exist. While all-sky spectroscopic surveys have observed many massive stars, the machine learning pipelines that produce the effective temperatures, surface gravities, and chemical compositions that would allow us to accurately classify our sample do not cover the parameter regime in which massive stars reside \citep[e.g.][]{garciaperez16}. As a result, we must use the heterogeneous classification data available on SIMBAD. For each star, we apply a decision tree that results in the star receiving a single label. Figure \ref{fig:classify} shows a flowchart that summarizes our labeling scheme. Note that this process highly tailored to this dataset, and some branches in the decision tree serve only to accurately label very small numbers of stars with unique spectral types (e.g., spectroscopically peculiar stars or X-ray binaries). Deriving labels for known massive stars using existing sources is not trivial, and our labelling scheme would be entirely different if a large sample of massive stars with well-measured temperatures, surface gravities, and chemical abundances were available.

We first use the common name and \Gaia ${\tt source\_id}$ of the star to determine if the star belongs to the catalog of confirmed Luminous Blue Variables (LBVs) presented in \citet{richardson18}. Non-LBVs are classifed as WR stars if ``W'' is in the spectral type, or the SIMBAD ${\tt OType}$ is ``*WR''. Non-WRs with ``K'' or ``M'' in their spectral type are classified as either Red Supergiants (RSGs) or ``C/S/Giant'' if their SIMBAD ${\tt SP\_Type}$ contains ``III'' --- we keep all such low-mass contaminants in our sample as distinguishing between RSGs and luminous low-mass giants is still a difficult problem \citep{massey09,yang19,neugent20}. The resulting sample of RSGs is pure; of the five RSGs that don't have luminosity class I, all of them are luminosity class Ia-II, Ib-II, or Iab-II, which are consistent with bona fide RSGs \citep{levesque05}. This is a good test of the \Gaia DR2 parallaxes and \citet{bailerjones18} distances, as cool subgiants and dwarfs with luminosity class IV or V would have been erroneously classified as RSGs using our criteria had they been included in our sample due to inaccurate distance and $M_G$ measurements.

Non-RSGs with ``F'' or ``G'' in their spectral type are classified as Yellow Supergiants (YSGs). However, this includes a number of blue stars. Further inspection of these stars reveals a number of objects whose \texttt{MK\_Spectral\_Type} field contradictorily indicates these are hot stars, with spectral type O, B, or A. As these stars have \Gaia photometry consistent with hot stars, we classify them as such (see below). Eight low-mass yellow stars are also included in our sample. Stars with ``III'' or ``V'' in their spectral types are classified as Yellow Dwarfs. While luminosity class III formally denotes giant stars, only one yellow giant is in the sample, and so we assign it the Yellow Dwarf label. We note that for extragalactic samples, foreground dwarfs can usually be filtered based on proper motions, while dwarfs belonging to the stellar population under study can be excluded just based on their apparent magnitudes. Nonetheless, we retain this class to avoid confusion between these stars and true YSGs. All YSGs that aren't hot stars or dwarfs keep their YSG label. 

Of the objects that have not yet been classified, stars with ``[e]'' in their spectral type are classified as OB[e] stars, while non-OB[e] stars with spectral types including an ``e'' (without brackets, and without ``pec'' in their spectral type) are classified as OBAe stars. If the star is not yet classified and O, B, or A are in the spectral type with no additional information, they are classified as generic OBA stars. OBA stars with ``III'' or ``IV'' in their spectral type are classified as evolved OBA stars, stars with ``V'' in their spectral type are classified as OBA main sequence stars, and finally, stars with ``I'' in their spectral type are labeled as OBA supergiants. All stars that have not been assigned a label at this stage are either C/S stars (which are assigned the C/S/Giant class), stars labeled only as Variables in SIMBAD (e.g., LPVs, semi-regular variables, or just variables, without other spectral information) which are assigned the Miscellaneous Variable classification, or stars with no identifying information/no confirmed designation (e.g., the SIMBAD ${\tt OType}$ is ``Star'' or contains ``Candidate'') which are classified as Unknown/Candidate. 

Finally, we include an ``Is Binary'' flag for all stars, which is 1 for stars classified as Eclipsing or Spectroscopic Binaries, High Mass X-ray Binaries, or Ellipsoidal Variables, or if they have a compound spectral type (e.g., WN8 + O6V),\footnote{This does not include stars with the ``OB+'' spectral type, which is an outdated class that describes OB stars with weaker absorption lines that would now be classified as OB supergiants.} and 0 otherwise; 102 stars are flagged as binaries. This flag is separate from the labeling process shown in in Figure \ref{fig:classify}. Because photometry of binary systems can be misleading \citep{neugent18b}, and binary systems exhibit a broad range of variability that is not intrinsic to the individual components, we exclude these stars from our classifier. 

\begin{figure}[ht!]
\plotone{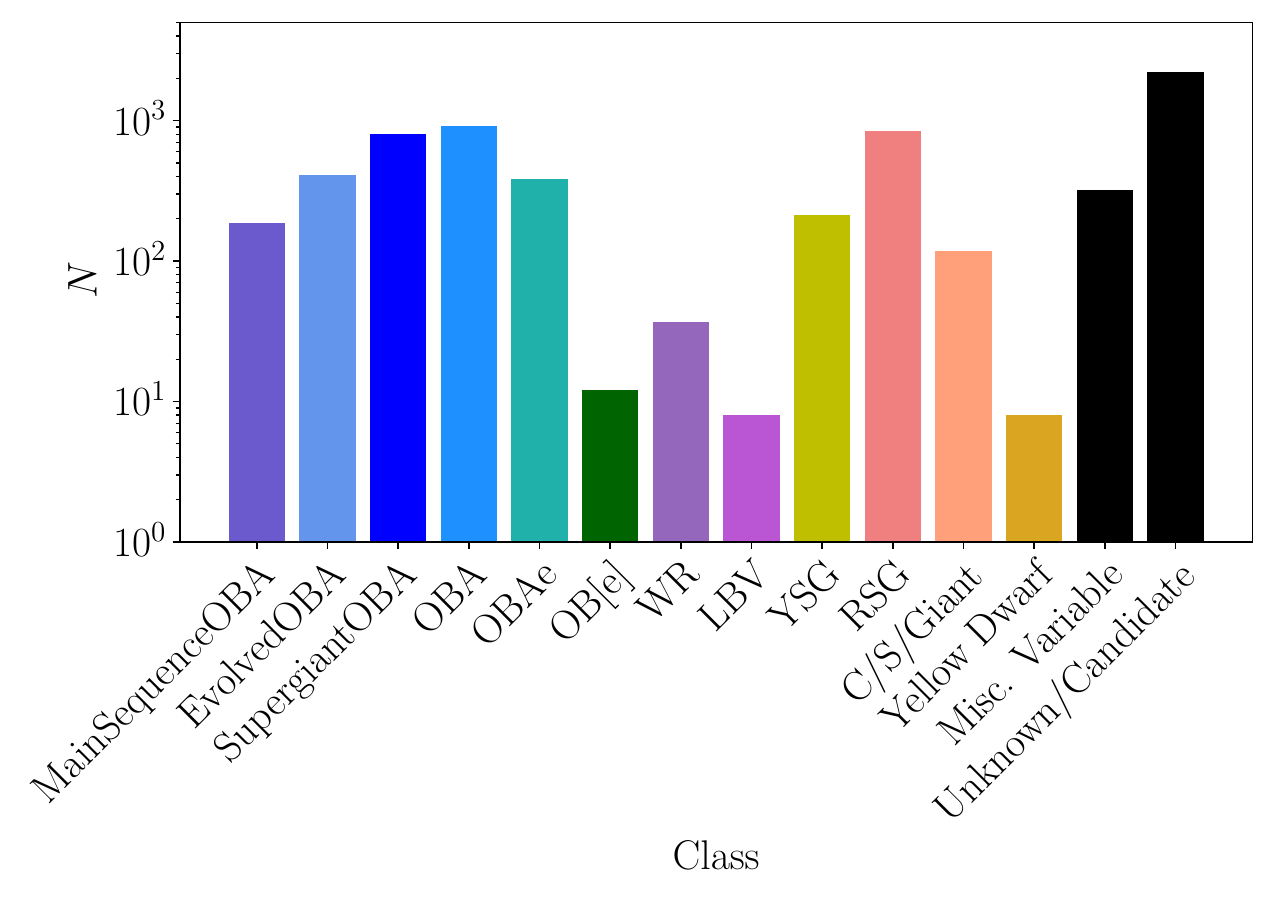}
\caption{The makeup of our sample of massive stars. Note that the sample is dominated by OBA stars and cool supergiants. Non-OBAe emission line stars --- OB[e] stars, WRs, and LBVs --- are the rarest massive stars in our sample, despite being stars of great scientific interest. For readability, we have used a logarithmic y-axis to display our sample statistics. Note that in practice, differences in the number of stars per class are much larger than they might appear here.}\label{fig:sample_makeup}
\end{figure}

Figure \ref{fig:sample_makeup} shows the makeup of our sample. Approximately 30\% of our sample (2550 stars) belong to the Miscellaneous Variable and Unknown/Candidate classes, which we do not use to train our classifier; instead, we use the classifier to assign tentative classifications in \S\ref{sec:svm}. The rest of the sample is dominated by luminous OBA stars and cool supergiants, with very few LBVs, OB[e] stars, and WR. This is unsurprising given these stars' high luminosity in the \Gaia bandpass, and the expected lifetimes of these evolutionary phases relative to the lifetimes of exotic emission line objects \citep{ekstrom12}. The imbalances in available training data across different classes, along with the extreme sparsity of training data in the rare classes, will impact the performance of the classifier if not properly addressed. \citep{chawla10}. We discuss this issue for this particular sample in \S\ref{subsec:classifier_selection}.

\begin{figure*}[ht!]
\plottwo{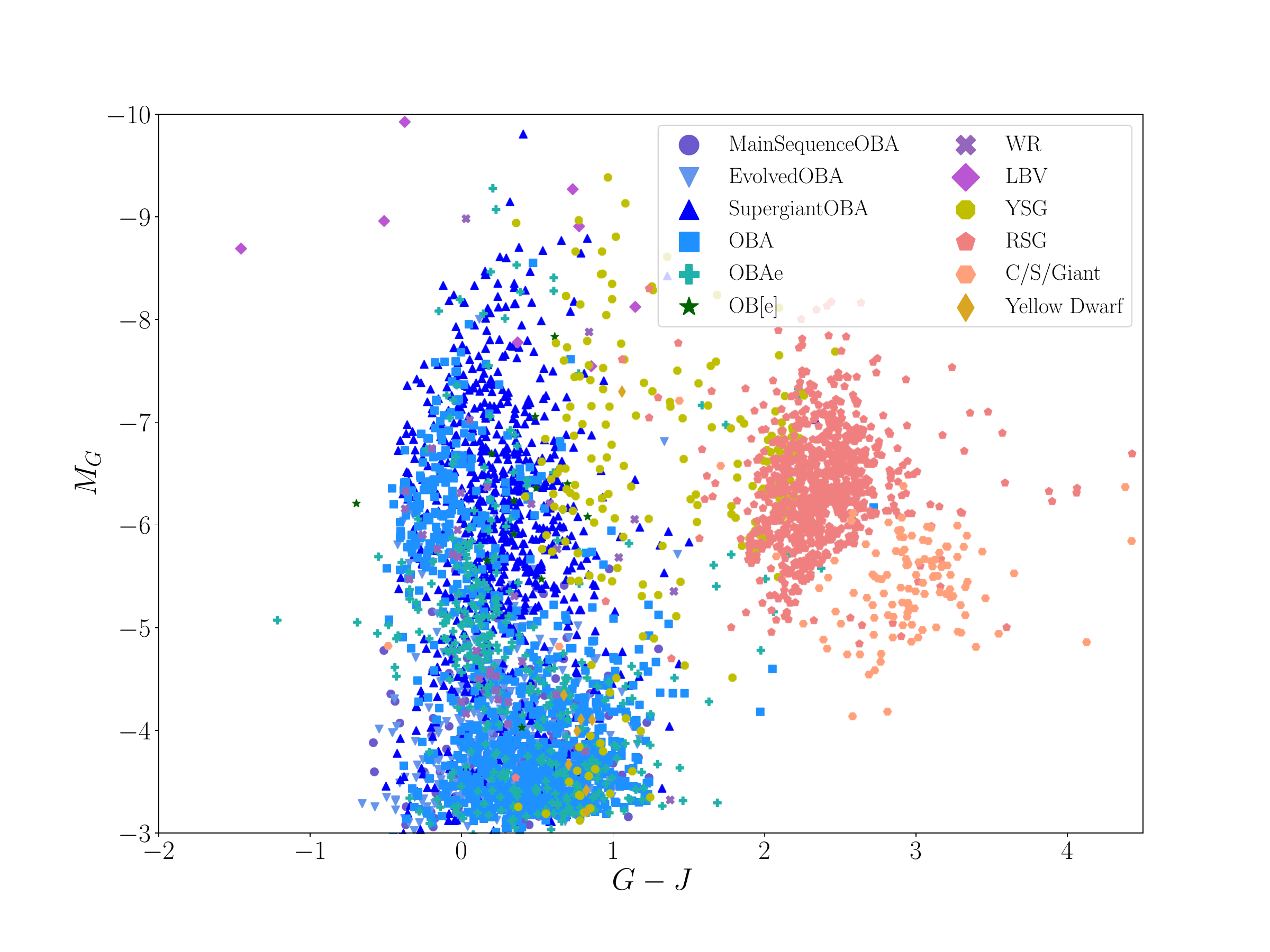}{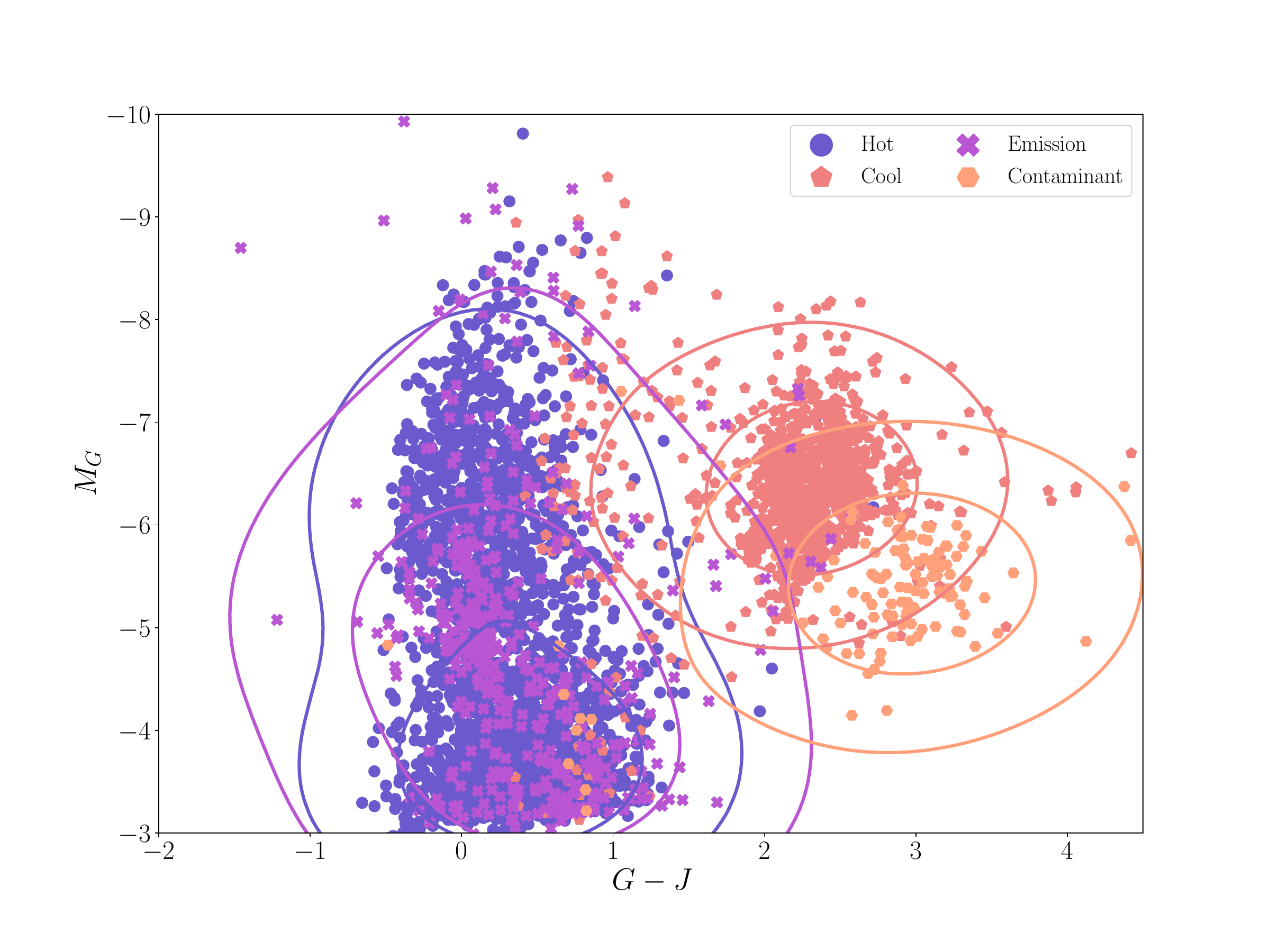}
\caption{$M_G$ vs. $G-J$ for putative massive stars. {\it Left}: Stars are colored by their label. While we also use shapes to distinguish between stars in different classes, this illustrates a key difficulty faced by our classifier: the classes have significant overlap with each other in the CMD. For example, the coolest/warmest YSGs have identical optical photometry to RSGs/OBA supergiants respectively, while the different classes of hot stars are impossible to distinguish from one another by eye. {\it Right}: Stars are colored by their coarse label. Contours for each coarse class correspond to 0.5 and 0.1 times the maximum value of a kernel density estimate of the distribution of each class in the CMD. Even in the coarse labels, the contours for hot stars and emission line stars are nearly identical.}\label{fig:labelcmd}
\end{figure*}

The left panel of Figure \ref{fig:labelcmd} shows the $M_G$ vs. $G-J$ CMD for all stars in our sample that aren't labeled as miscellaneous variables or unknown/candidate, colored by their label. $G-J$ correlates reasonably well with effective temperature in main sequence stars \citep{davenport2018}, and in this case is especially useful for distinguishing from the near vertical main sequence/blue supergiants and the significantly cooler yellow and red supergiants. From this plot, it is clear that many stars are misclassified in SIMBAD (the worst example is one particular red star classified as an OBA star), reducing the effectiveness of any machine learning algorithm, and propagating biases into the results. Yellow supergiants are especially prone to this problem: 81/212 YSGs in the sample (38\%) have $G-J<1$, consistent with the optical colors of much hotter stars. Indeed, this problem would have been worse had we not corrected for the presence of OBA stars in the initial sample of YSGs. This issue may originate from bad distance estimates for individual stars (which explains why our sample includes F and G dwarfs), bad estimates of reddening (given our usage of monochromatic extinction coefficients), previously unidentified variability, or the fact that many of these spectral types were determined via stellar spectra taken on photographic plates \citep[for example, many spectral types for stars in the LMC come from][]{ardeberg72}. We expect this issue to propagate into our results, increasing the confusion between YSGs and hot stars.\footnote{It would certainly be possible to tailor our dataset by removing the ``worst'' stars, thus cleaning up the boundaries between classes. However, by doing this we would be making several assumptions about where different classes of stars reside in our feature space, with no way of knowing whether these enforced boundaries actually divide stars in physically different evolutionary states. Indeed, we {\it expect} the boundaries between classes to be fuzzy, because our discrete labels are an approximation of a continuum of evolutionary states --- a fact that we have otherwise swept under the rug. That said, a significant amount of the overlap between classes {\it is} due to the poor quality of existing labels. Instead of trying to guess which stars are poorly labelled, and which ones truly reside in the overlap between classes, we instead wish to see how the quality of existing labels impacts the performance of our classifier.}

Because we expect objects in some classes --- especially those with an evolutionary link such as main sequence, evolved, and supergiant OBA stars --- to appear similar in the training data set, we also assign all stars a coarse label: all classes of OBA stars excluding OB[e] and OBAe are labeled ``Hot''; RSGs and YSGs are labeled ``Cool''; WRs, LBVs, and both OB[e] and OBAe stars are labeled ``Emission'' (EM for short); C/S/Giant stars and Yellow Dwarfs are labeled ``Contaminant''; and miscellaneous variables and unknown/candidates are labeled ``Unknown/Candidate.'' The results of this labeling scheme are summarized in Table \ref{tab:classes} which shows the number of stars with a given refined class that are assigned a particular coarse label. The right panel of Figure \ref{fig:labelcmd} shows the same CMD, with points colored by their coarse label. This leads to some improvement: each coarse class lies in the approximate region of the CMD that one would expect. Regardless, it is evident that selecting any one of these classes solely from this optical photometry would be difficult: the ``cool'' class has significant overlap with the ``hot'' class --- largely driven by the YSGs --- emission line stars can be found at a range of $G_{BP}-G_{RP}$ colors, and there is significant overlap between low-mass contaminants that will end their lives as white dwarfs and true massive stars that will end their lives in supernovae explosions. This point is emphasized by the contours, which correspond to 0.5 and 0.1 times the maximum value of a kernel density estimate of the distribution of each class, which replaces each point with a kernel function (in this case a two-dimensional Gaussian centered on the point), and sums the kernels to estimate the underlying distribution. We do this using the {\tt KernelDensity} estimator from {\tt sklearn}, and use a similar cross-validation scheme described below to find a suitable bandwidth for the kernel (i.e., the width parameter of the Gaussian). 

We use these coarse labels to train a second classifier. While these coarse labels lose some specificity, each coarse class contains more stars, hopefully increasing the performance of a classifier trained on these labels. Furthermore, they still retain physical information while increasing the number of stars in each class: the ``cool'' label contains stars with convective envelopes, while ``hot'' stars contain radiative envelopes. Meanwhile, emission line stars are notable for their variability. It is our hope that this second classifier will still address two of our stated goals: to identify emission line stars, and to reject contaminating low mass stars.

\begin{deluxetable*}{lcccc}
\tabletypesize{\scriptsize}
\tablecaption{Number of stars in a class that are assigned a given coarse label, not including the Miscellanous Variable or Unknown/Candidate labels.\label{tab:classes}}
\tablehead{ 
\colhead{} & \multicolumn{4}{c}{Coarse Label} \\
\cmidrule{2-5} \vspace{-15pt} \\
\colhead{Refined Label} & \colhead{Hot} & \colhead{Emission} & \colhead{Cool} & \colhead{Contaminant}
} 
\startdata
    Main Sequence OBA & 187 & ~    & ~        & ~           \\
    Evolved OBA       & 409 & ~    & ~        & ~           \\
    Supergiant OBA    & 798 & ~    & ~        & ~           \\
    OBA               & 915 & ~    & ~        & ~           \\
    OBAe              & ~   & 383  & ~        & ~           \\
    OB[e]             & ~   & 12   & ~        & ~           \\
    WR                & ~   & 37   & ~        & ~           \\
    LBV               & ~   & 8    & ~        & ~           \\
    YSG               & ~   & ~    & 212      & ~           \\
    RSG               & ~   & ~    & 847      & ~           \\
    C/S/Giant         & ~   & ~    & ~        & 118         \\
    Yellow Dwarf      & ~   & ~    & ~        & 8           \\
    \cmidrule{1-5}
    Total             & 2309 & 440 & 1059 & 126 \\
\enddata
\end{deluxetable*}

\section{WISE Lightcurves}

Variability in evolved massive stars has been well-characterized at timescales from minutes to decades \citep[e.g.][]{conroy18,dornwallenstein19,soraisam20}. In a study of massive stars in the Whirlpool Galaxy (M51), \citet{conroy18} found that almost half of the stars brighter than $M_I = -7$ were variable, with red stars nearing a variability fraction of 1. Both red and extremely luminous blue stars exhibited quite high amplitude ($\Delta I \geq 0.3$) variability. For spectral energy distributions (SEDs) dominated by purely stellar light, mid-infrared (MIR) flux measurements (and thus variability) is sensitive to (variations in) the bolometric luminosity. However, for stars with significant circumstellar dust components in their SEDs, MIR variability is correlated with both intrinsic bolometric variability, and with dust creation/destruction processes in the circumstellar medium \citep[for example, in RSGs where it is correlated with the mass loss rate, e.g.][]{yang18}.

The WISE mission provides lightcurves from stars in all parts of the sky, observed over a $\sim$7 year baseline. Due to the scanning law adopted by WISE \citep{wright10}, most stars not on the ecliptic poles are visited approximately every $\sim180$ days. All stars have a $\sim3$-year data gap from when WISE was placed in hibernation in February 2011 and when it was reactivated in December 2013. WISE initially observed simultaneously in four filters during its primary mission: $W1$ (3.4 $\mu$m), $W2$ (4.6 $\mu$m), $W3$ (12 $\mu$m), and $W4$ (22 $\mu$m). However, it was reduced to using only the two bluest bands in its post-cryogenic survey mode called ``NEOWISE''.
The time, duration, and number of individual observations during each $\sim180$ day visit depends on spatial geometry of the WISE scanning program, i.e. stars closer to the ecliptic poles have longer duration visits (often exceeding a week) with many epochs per visit, while star near the equator have very short visits (typically a couple days) with only a few epochs per visit. Because WISE lightcurves possess such non-uniform cadence, extracting detailed physics for most individual stars is difficult. However, the WISE lightcurves place {\it fantastic} constraints on MIR variability amplitudes on longer timescales, especially for evolved massive stars whose highest-amplitude variability occurs over $\sim$year timescales. Such amplitude and timescale estimates are related to the physical parameters of the star, potentially aiding in classification.

For every star selected in \S\ref{sec:sample} we queried the Single-Exposure (``L1b'') source databases for all phases of the WISE mission, including the original 4-band, partial cryogenic 3-band, and post-cryogenic 2-band NEOWISE tables. We used {\tt astroquery} to pull data in the region within 3 arcsec of the known source location. To ensure high quality data for all recovered epochs, we require the photometric quality flag to be {\tt PH\_QUAL=A}, the contamination flag to be {\tt CC\_FLAGS=00}, the number of deblended sources flag to be {\tt NB=1}, and the PSF photometry fit quality (defined as the reduced $\chi^2$) in $W1$ to be {\tt w1rchi2$<$5}.

\begin{figure*}[ht!]
\includegraphics[trim=0mm 0mm 0mm 0mm, clip, width=\textwidth]{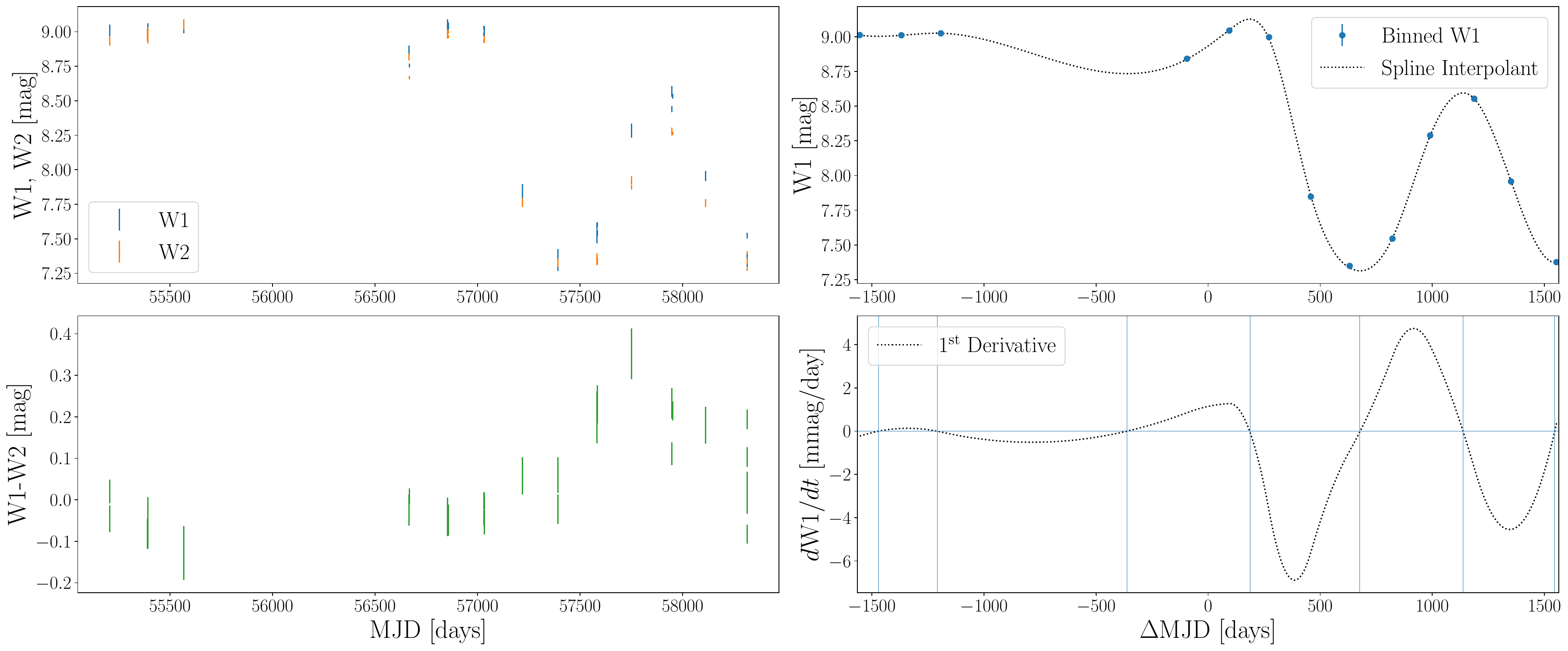}
\caption{Example lightcurve for WISE J000536.97+432405.0. {\it Top left}: Raw lightcurve, with $W1$ points plotted as blue errorbars, and $W2$ points plotted in orange. {\it Bottom left}: Variability in $W1-W2$ plotted as green errorbars. {\it Top right}: Binned $W1$ lightcurve. Blue points are binned data (errorbars are smaller than the points). Black dotted line is the B-spline interpolation. Time has been adjusted so the lightcurve is centered on $t=0$. {\it Bottom right}: First derivative of the interpolant. Vertical blue lines show the times where the derivative crosses zero, indicated by the horizontal blue line.}\label{fig:ex_lc}
\end{figure*}

Only two of our stars did not have usable data from WISE: WISE J074911.48-102000.2 (HD 63554) has no lightcurve available online, and WISE J050128.62-701120.2, which does not have any corresponding object nearby on SIMBAD. When calculating each of the variability metrics below, we instead record a value of \texttt{NaN} (i.e., missing data). For the remaining stars, we ignore the $W3$ and $W4$ data here due to the lower signal to noise and significantly shorter observing baselines due to the loss of cryogenic observations after the original WISE mission. As WISE observes simultaneously in all bands, we can construct $W1-W2$ lightcurves without any interpolation, and simply subtract the $W2$ data from the $W1$ data to obtain the $W1-W2$ color curve. The left panels of Figure \ref{fig:ex_lc} show an example set of lightcurves for WISE J000536.97+432405.0 (=HD 73), a B1.5IV star that illustrates the typical observing cadence and variability of a bright star in our sample.

\subsection{Variability Metrics}\label{sec:variabilitymetrics}

\subsubsection{Amplitude}

For each of the three lightcurves of each object, we wish to extract simple metrics that describe the amplitude and timescale of variability. We choose $\chi^2$ about the median defined as 
\begin{equation}
    \chi^2 = \sum \big(\frac{M_i-\Tilde{M}}{\sigma_i}\big)^2
\end{equation}
and the reduced-$\chi^2$
\begin{equation}
    \chi^2_{red} = \chi^2/(N-1)
\end{equation}
Where $M_i$ is a magnitude measurement, $\Tilde{M}$ is the median of the lightcurve, $\sigma_i$ is the corresponding error on the data point, and $N$ is the number of points in the lightcurve. We also calculate the Median Absolute Deviation (MAD), and Error-Weighted MAD (EWM):
\begin{align}
    MAD &= \mathrm{Median}(|M_i-\Tilde{M}|) \\
    EWM &= \mathrm{Median}(|M_i-\Tilde{M}|/\sigma_i)
\end{align}
If the filtered and cleaned lightcurve only contains one good measurement (or no good measurements), we automatically give it $\chi^2 = \chi^2_{red} = {\rm MAD} = {\rm EWM} = {\tt NaN}$. We describe our method for treating missing data below.

\begin{figure}[ht!]
\plotone{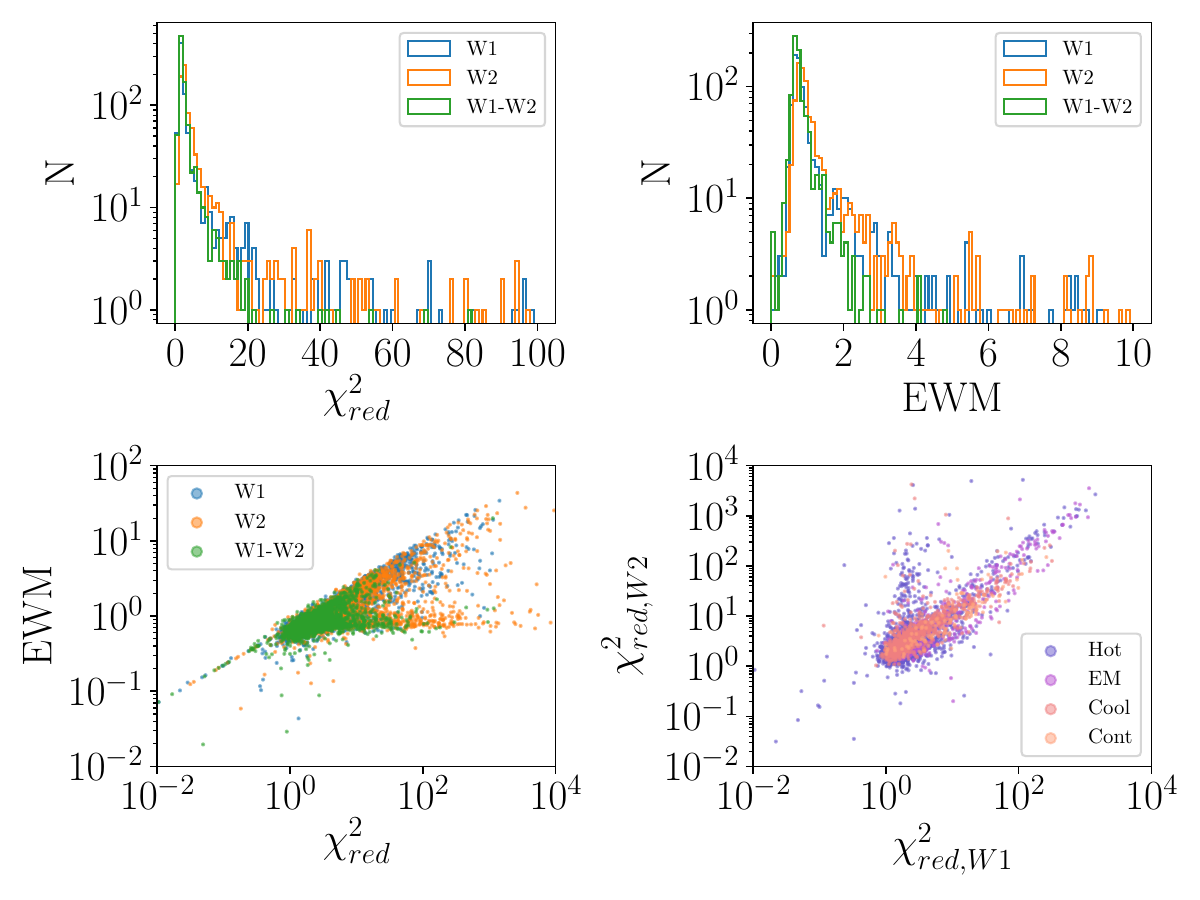}
\caption{{\it Top row}: Distribution of derived $\chi^2_{red}$ (left) and $EWM$ (right) values in the $W1$ (blue), $W2$ (orange), and $W1-W2$ (green) lightcurves. {\it Bottom row}: Scatter plots comparing different amplitude metrics. The left panel shows $EWM$ vs. $\chi^2_{red}$ in the $W1$, $W2$, and $W1-W2$ lightcurves (using the same color-coding). The right panel shows $\chi^2_{red}$ in $W2$ vs. $\chi^2_{red}$ in $W1$, with each point colored by its coarse label.}\label{fig:chi_ewm}
\end{figure}

The top panels of Figure \ref{fig:chi_ewm} show the distributions of $\chi^2_{red}$ and EWM derived for our sample. Values from $W1$ lightcurves are in blue, $W2$ in orange, and $W1-W2$ in green. The bottom left panel shows a scatter plot of $\chi^2_{red}$ vs. EWM for all three lightcurves. While the two measures correlate reasonably well with each other, there is a branch of stars whose lightcurves have high $\chi^2_{red}$ and low EWM; because the EWM is robust to outliers, $\chi^2_{red}$ is an effective probe of lightcurves with sudden brightening/fading events, while EWM is an effective selector for lightcurves that display consistent variability. The bottom right panel shows a scatter plot of $\chi^2_{red}$ in $W1$ vs. in $W2$, with each point colored by its coarse label. A distinct branch of stars that are much more variable in $W2$ than $W1$ is clearly evident; oddly, the distributions of classes, EWM, and broadband colors in this branch are consistent with the whole sample, and no similar branch exists in the measured EWM values. 

Visual inspection of the lightcurves of stars with $\chi^2_{red} < 10$ in $W1$ and $\chi^2_{red} > 100$ in $W2$ shows that these stars appear to have higher signal to noise $W2$ measurements than $W1$, and have one observation during which the star apparently becomes considerably redder, achieving $W1-W2$ values as high as $\sim4$. Examining the times at which these extreme reddening events occur shows a preference for times during the cryogenic WISE survey, implying that this behavior is likely instrumental in origin, despite our filtering using the provided quality flags. Nonetheless, we include $\chi^2_{red}$ as it does not map perfectly onto $EWM$, and only 98 stars fall in this regime. We do not yet know whether $\chi^2_{red}$, $EWM$, both metrics, or neither are useful features for classification, so we keep both with the intent of exploring their importance below.

\subsubsection{Timescale}

Many methods exist for estimating dominant timescales in lightcurves. \citet{conroy18} use the Lomb-Scargle Periodogram \citep{lomb76,scargle82} to search for periodic variables. However, this approach suffers from numerous, well-known issues (including accurate period recovery at low signal to noise), and false peaks can easily be mistaken for real timescales, especially in highly-irregularly sampled data, as is the case for the WISE lightcurves. \citet{soraisam20} use a Gaussian process interpolation scheme coupled with a wavelet analysis to estimate timescales in massive stars in M31 observed by the Palomar Transient Factory (PTF). However, with so few data points, we found it difficult to obtain a reliable fit with a Gaussian process, and even when the fit was successful, the resulting interpolant had a large standard deviation in between WISE visits. The resulting measurements of the characteristic timescale was more reflective of the kernel used.

Instead, we turn to a a spline-based interpolation method, which is analogous to certain Gaussian process methods \citep{kimeldorf70}. We first subtract half of the sum of the times of the first and last available observations, so that the lightcurve is centered at $t=0$. We then bin the observations in each visit. Visits are defined as sets of points separated in time by less than a defined threshold. Due to the WISE scanning law, some stars near the ecliptic poles have visits separated by less than the typical $\sim180$ days. Therefore, we adopt 50 days as the threshold for visits. Two stars in our sample are close enough to the ecliptic pole to be observed nearly continuously such that we erroneously record two ``visits'': one each during the cryogenic and post-cryogenic surveys. However, neither star is strongly variable and thus this small edge case does not substantially impact our subsequent analyses.

For all observations in a given visit, we calculate the mean time and $W1$/$W2$/$W1-W2$ measurement. We use \texttt{scipy.interpolate.splrep} in Python to find the 3$^{\rm rd}$-order basis spline (a.k.a. B-spline, which performs a spline fit using spline basis functions, \citealt{deboor78}) representation of the binned lightcurves, adopting a smoothing factor \texttt{s=10}. This returns the knots, B-spline coefficients, and degree of the spline. By definition, 3$^{\rm rd}$-order splines are differentiable, so we use \texttt{scipy.interpolate.splev} to evaluate the first derivative of the spline interpolant, and find the times when the derivative changes sign --- i.e., when the lightcurve reaches a maximum or minimum. As metrics of the characteristic timescale of the lightcurve, we calculate the frequency of zero-crossings of the first derivative of the spline interpolant, $\nu_0$ --- calculated as the number of times the derivative passes through zero, divided by the time baseline of the lightcurve --- $\langle \Delta t \rangle$, the mean of the differences between successive zero-crossings, and the standard deviation of the differences between successive zero-crossings, $\sigma_{\Delta t}$. For stars with fewer than four visits, we automatically assign $\nu_0 = \langle \Delta t \rangle = \sigma_{\Delta t} = {\tt NaN}$. The right panels of Figure \ref{fig:ex_lc} show this process on the $W1$ lightcurve plotted in the left panels. The blue points are the binned $W1$ measurements (the errors are smaller than the size of the points), and the dotted black lines are the spline interpolant (top right) and corresponding derivative (bottom right). 

While this is a simple method that yields multiple estimates of variability timescale, it is important to note that it is dependent on both the variability amplitude and the sampling. For example, a non-variable object whose lightcurve is poorly sampled may appear to be variable due to measurement noise (which does not have a characteristic timescale), and the derived timescale from this method will thus be more reflective of the sampling than anything else. Thankfully, in many cases such a variable would have a low $EWM$ value. However, it is possible that a star may enter a period of low-amplitude variability resulting in false zero-crossings of the first derivative of the spline interpolant (e.g., the first three WISE visits in lightcurve in Figure \ref{fig:ex_lc}; it is possible that the first few zero crossings in the bottom right panel may not be real.) These systematics are difficult to work around in sparsely sampled lightcurves, and are an important caveat to keep in mind.

\section{Machine Learning}\label{sec:svm}

\subsection{Classifier Selection}\label{subsec:classifier_selection}

The problem of classification based on broad-band photometry has a rich history in the literature. With the advent of large surveys like the Sloan Digital Sky Survey (SDSS, \citealt{york00}), optical data could be coupled with space-based MIR data to find the stellar locus in a 10-dimensional color-space \citep{davenport14}. Recent efforts to separate stars from quasars, or perform a regression on effective temperature with machine learning on photometric data have been successful \citep{makhija19,bai19}; however, these studies are often focused on main sequence, low mass stars. This is an understandable choice given the rarity of evolved, high mass stars, the absence of reliable distances to calculate luminosities from which to select putative massive stars, and the fact that follow-up spectroscopy is necessary in order to confirm a star's membership in many important classes.

With the advent of \Gaia DR2, luminosities can be easily determined, and putative massive stars can be confirmed, as we do in \S\ref{sec:sample}. We wish to train an algorithm that takes as input the broadband photometry and variability metrics derived for our sample, and outputs spectral type classifications. Many machine learning classifiers exist; of these, we wish to choose a flexible model with well-understood mathematics, while avoiding techniques like neural networks that can be difficult to interpret. Of the classifiers available in the \texttt{sklearn} package, we decided to test a Random Forest (RF) classifier \citep{breiman01} --- which consists of a collection of decision trees trained on random subsets of samples and features --- a Support Vector Machine (SVM) classifier \citep{cortes95} --- which identifies hyperplanes in the feature space that separate different classes --- and a Gaussian process (GP) classifier --- which models the function determining the probability of a star being a given class at a location in the feature space as a multidimensional Gaussian distribution whose properties are determined entirely by a covariance function (a.k.a. a kernel function), coupled with a linking function (usually the logit function) to make discrete class predictions \citep{rasmussen06}. We refer the reader to these publications, as well as to the \texttt{sklearn}  documentation\footnote{\url{https://scikit-learn.org/stable/index.html}} for the mathematics and implementation details of each classifier. In the multi-class case, a collection of classifiers are trained on each possible pair of classes (one-versus-one or ``ovo''), generating a total of $N_{classes}(N_{classes}-1)/2$ classifiers where $N_{classes}$ is the number of classes. Labels are assigned to test samples by allowing each classifier to vote, and the label with the most votes is chosen \citep{knerr90}.

\begin{deluxetable}{lr}
\tabletypesize{\footnotesize}
\tablecaption{List of features passed to our machine learning classifiers, as well as clarifying definitions where relevant. WISE photometry used to calculate colors and magnitudes is from the ALLWISE data release \citep{cutri13}. All variability metrics are calculated from the WISE $W1$, $W2$, and $W1-W2$ lightcurves. \label{tab:features}}
\tablehead{\colhead{Feature}  & \colhead{\hfill Definition}} 
\startdata
\multicolumn{2}{c}{Colors \& Magnitudes} \\
\cmidrule{1-2}
$M_G$ & Absolute magnitude in \Gaia $G$ band. \\
$G-J$ &  From \Gaia and 2MASS photometry.  \\
$J-H$ &    From 2MASS photometry. \\
$H-K_s$ &   From 2MASS photometry.   \\
$K_s-W1$ &   From 2MASS and WISE photometry.   \\
$W1-W2$ &   From WISE photometry.   \\
$W2-W3$ &   From WISE photometry.  \\
$W3-W4$ &   From WISE photometry.   \\
$M_{W1}$ & Absolute magnitude in WISE $W1$ band. \\
\cmidrule{1-2}
\multicolumn{2}{c}{Variability Metrics} \\
\cmidrule{1-2}
$\log\chi^2_{red}$ &  Log of the reduced $\chi^2$.   \\
$\log$ EWM &   Log of the error-weighted Median Absolute Deviation.   \\
$\nu_{0}$ &  Frequency of zero-crossings of the first derivative of the \\
& spline interpolant. \\
$\log\langle \Delta t \rangle$ &  Log of the average time between zero-crossings.   \\
$\log\sigma_{\Delta t}$ &  Log of the standard deviation of zero-crossing times. \\
\enddata
\end{deluxetable}

Each type of classifier has a number of hyperparameters that affect the performance of the classifier. For the RF classifier, \texttt{n\_estimators} specifies the number of trees in the forest, \texttt{max\_depth} specifies how many branches each decision tree in the forest can have, and \texttt{max\_features} specifies the maximum number of features each tree is trained on. We also set \texttt{class\_weight=balanced}, which weighs samples when fitting to account for the different frequencies of each class in the data. 

For the SVM classifier (SVC), $C$ is a regularization parameter that governs the tradeoff between maximizing the margin and misclassifications in the training set. Higher values of $C$ will force the SVC to correctly classify every point, resulting in poor generalization (i.e. overfitting). The SVC requires that the distance between two points in the feature space is defined as the inner product of two vectors in the feature space, $\langle \Vec{X_i},\Vec{X_j}\rangle$. Because the boundaries between classes in our sample are not guaranteed to be linear, one can project the samples into a much higher dimension space via a mapping function, $\Phi$, where distances between two vectors in this space are calculated as $\langle \Vec{Z_i},\Vec{Z_j}\rangle = \langle \Phi(\Vec{X_i}),\Phi(\Vec{X_j})\rangle$. In reality, the transformed feature space can be incredibly high dimensional, and explicitly mapping the data into this high-dimensional space is computationally inefficient. Instead, we can adopt a kernel function, $K$ that defines distances in the higher dimensional space, e.g., $K(\Vec{X_i},\Vec{X_j}) = \langle \Vec{Z_i},\Vec{Z_j}\rangle$. Because the kernel only takes the measured features as input and outputs a number, using a kernel function implicitly maps the input feature space into a high dimensional space without specifying $\Phi$. 

Common choices of the kernel function include a linear kernel (i.e., the Euclidean distance between individual samples in the feature space) and the ``radial basis function'' (RBF) kernel:
\begin{equation}
    K(\Vec{X_i},\Vec{X_j}) = e^{-\gamma||\Vec{X_i}-\Vec{X_j}||^2}
\end{equation}
where $\gamma$ governs the influence of the kernel function; lower values result in increasingly linear boundaries, while high values result in the decision function being entirely depended on individual points, creating small islands of a given class centered on each training point. The advantage of non-linear models like a SVM with a RBF kernel over linear methods is that the decision boundaries can be much more flexible; the tradeoff is that the contribution of individual features to the classifier cannot be easily calculated without the unknown function $\Phi$ (see \S\ref{subsubsec:featureimportance}). The ``optimum'' kernel and hyperparameters are chosen via a cross-validation strategy described below. Finally, we also set \texttt{class\_weight=balanced} for the SVC, which automatically sets the value $C$ for class $i$ to $C N_{\rm samples}/(N_{\rm classes}N_i)$ where $N_{\rm samples}$ is the size of the sample, $N_{\rm classes}$ is the number of classes, and $N_i$ is the number of objects in the sample belonging to the class. This serves to weight rarer classes more heavily. The GP classifier's only hyperparameter is a choice of kernel, which defines the covariance function of the Gaussian process.

To fit our classifiers, we first remove all stars labeled as miscellaneous variables or unknown/candidates, as well as known binaries, and one star with bad $J$ photometry, the Be star HD 53032. For features, we use the intrinsic calculated value of $M_G$, as well as (uncorrected for extinction) $G-J$, $J-H$, $H-K_s$, $K_s-W1$, $W1-W2$, $W3-W4$, and $\chi^2_{red}$, EWM, $\nu_0$, $\langle \Delta t \rangle$, and $\sigma_{\Delta t}$ in all three lightcurves --- we indicate the WISE band or color that each variability metric corresponds to with a subscript hereafter. The input features and a brief description where relevant are listed in Table \ref{tab:features}. Because $\chi^2_{red}$, EWM, $\langle \Delta t \rangle$, and $\sigma_{\Delta t}$ have significant dynamic range, we use the base-10 logarithm of these features. Features values as well as labels for each star in our sample are given in Table \ref{tab:feature_values}. We note that only a subset of the features from Table \ref{tab:features} are listed here due to the number of features. The table in its entirety will be made available in a machine-readable format.

\centerwidetable
\begin{deluxetable*}{lccccchrr}
\tablefontsize{\scriptsize}
\tablecaption{Feature values and assigned labels for all stars in our sample, ordered by Right Ascension. Missing numbers are indicated with ``-''. \label{tab:feature_values}}
\tablehead{\colhead{Common Name} & \colhead{$M_G$ [mag]} & \colhead{$G-J$ [mag]} & \colhead{$W1-W2$ [mag]} & \colhead{$\log\chi^2_{red,W1}$} & \colhead{$\log\langle \Delta t \rangle_{W1}$ [d]} & \colhead{Label} & \colhead{Coarse Label}} 
\startdata
HD 236270 & $-3.54$ & $0.35$ & $0.14$ & $0.389$ & $nan$ & RSG & Cool \\ 
LS   I +64   10 & $-3.41$ & $0.72$ & $0.00$ & $nan$ & $nan$ & OBA & Hot \\ 
LS   I +60   69 & $-3.22$ & $0.88$ & $-0.03$ & $0.061$ & $nan$ & OBAe & EM \\ 
BD+62  2353 & $-4.27$ & $0.47$ & $-0.04$ & $0.110$ & $2.816$ & SupergiantOBA & Hot \\ 
HD     73 & $-3.29$ & $-0.66$ & $-0.05$ & $3.157$ & $2.701$ & EvolvedOBA & Hot \\ 
HD 240496 & $-3.84$ & $1.01$ & $0.01$ & $-0.064$ & $2.764$ & OBA & Hot \\ 
WISE J000559.28-790653.3 & $-5.33$ & $1.34$ & $-0.04$ & $0.072$ & $3.162$ & Unknown/Candidate & Unknown/Candidate \\ 
LS   I +59   30 & $-3.34$ & $0.67$ & $-0.04$ & $0.234$ & $nan$ & OBA & Hot \\ 
BD+57  2870 & $-4.53$ & $1.17$ & $0.00$ & $0.046$ & $3.135$ & SupergiantOBA & Hot \\ 
BD+62     1 & $-3.31$ & $0.90$ & $0.27$ & $1.622$ & $2.813$ & EvolvedOBA & Hot \\ 
\enddata
\tablecomments{This table is published in its entirety in the machine-readable format. A portion is shown here for guidance regarding its form and content. We note that only a subset of the features listed in Table \ref{tab:features} are shown here. All features are listed in the machine-readable version.}
\end{deluxetable*}

We now randomly split our sample into a training set with 70\% of the samples, and a test set with the remaining 30\%, using a stratification strategy to ensure the proportions of the classes in both sets are equal. The test set is withheld until we are ready to assess the performance of the chosen classifier. We then use \texttt{sklearn.preprocessing.StandardScaler} in Python to scale the training data such that each feature has 0 mean and unit variance. Because the data have missing values, we then use \texttt{sklearn.impute.IterativeImputer} which uses a Bayesian ridge regression to predict and replace missing values. 

To test the accuracy of the imputer, we select only the rows from the training set with no missing data. For each feature with missing data ($\log \chi^2_{red,W1-W2}$, $\log {\mathrm EWM}_{W1-W2}$, $\log \sigma_{\Delta t, W1}$, $\log \sigma_{\Delta t, W2}$, and $\log \sigma_{\Delta t, W1-W2}$) we randomly choose 200 objects, replace the value of the feature with \texttt{NaN} for only these objects, transform the data using the scaler and imputer, and calculate the fractional error between the true value and the imputed value. The returned fractional errors for each feature centered around 0, and had a low scatter with the exception of $\log {\mathrm EWM}_{W1-W2}$. However, this has little impact on the classifier, as only two objects in the actual training set have missing values for $\log {\mathrm EWM}_{W1-W2}$. We also repeated this procedure for coarse labels: we select 200 random objects with a given coarse label, replace a random feature from the list of features with missing data with \texttt{NaN} for each object, and again calculate the fractional error between the true and imputed values. We find that the imputer performs poorly on Cool and Contaminant stars. Given that all of the features with missing data are linked to variability, a significant fraction of red supergiants display high amplitude variability \citep{conroy18}, the MIR variability of AGB stars (which make up the bulk of the Contaminants) is higher than RSGs in a given magnitude range \citep{yang18}, and our sample of Cool and Contaminant stars contains objects in quite different evolutionary states that nevertheless have similar colors and magnitudes, it is unsurprising that the imputer is unable to predict the variability properties of these stars. In \S\ref{subsubsec:featureimportance}, we will discuss the impact of these features on the overall performance of our classifier. 

For each classifier, we then initialize a corresponding \texttt{sklearn} classifier object (e.g., \texttt{sklearn.svm.SVC}). To settle on the best values for the hyperparameters we use \texttt{sklearn.model\_selection.GridSearchCV} to perform a cross-validation search on a grid of hyperparameters, using a stratified K-fold strategy with $k=5$ to ensure that each fold has a representative distribution of classes. For the RF, we search for \texttt{n\_estimators} between 10 and 150 in steps of 10, \texttt{max\_depth} between 10 and 100 in steps of 10, and allow \texttt{max\_features} to be either \texttt{sqrt}, \texttt{log2}, or \texttt{None} (where the maximum number of features individual trees are trained on is the square root of, base-2 logarithm of, or equal to the number of features, see the documentation for details). We also allow \texttt{max\_depth} to take on the default value (\texttt{None}), such that individual trees can be grown until each leaf only contains one sample. 

For the SVC, we search for values of $C$ on a logarithmic grid with 1 dex spacing between 0.01 and 100, and for the RBF kernel, we search for values of $\gamma$ on a similar grid between 0.01 and 10. Additionally, we allow $\gamma$ to be the default values of \texttt{1/n\_features} (where \texttt{n\_features} is the number of features). For the GP classifier, we only vary the kernel, as the \texttt{GaussianProcessClassifier} object automatically optimizes the kernel hyperparameters. We let the kernel be either linear, RBF, or the default (a special case of the RBF kernel with the length scale equal to 1). 

Each classifier object has a default method to score each set of hyperparameters, e.g. the accuracy of predicted labels compared to true labels. However, the classes in our training set are unbalanced (e.g. Figure \ref{fig:sample_makeup}), so inaccurately classifying every single LBV, for example, would have little impact on the overall accuracy of the classifier. To account for this, we instead use the {\it balanced accuracy} \citep{mosley13,guyon15}, which weighs each sample by the frequency of that sample's class in the training set. Other options for scoring criteria exist, including some that help maximize the classifier's precision such as the weighted $F_1$ score and Cohen's kappa \citep{cohen60}. We experimented with using these scores, and found that using the balanced accuracy minimizes misclassifications across all classes (reflected in the diagonal in the left panels of Figures \ref{fig:matrix_plots} and \ref{fig:matrix_plots_coarse}). Note that this choice implicitly selects a classifier that performs well across all classes, and is not optimized for specific classes. Future work will explore the possibility of tuning a classifier to find specific classes of rare stars. 

\begin{figure}[ht!]
\plotone{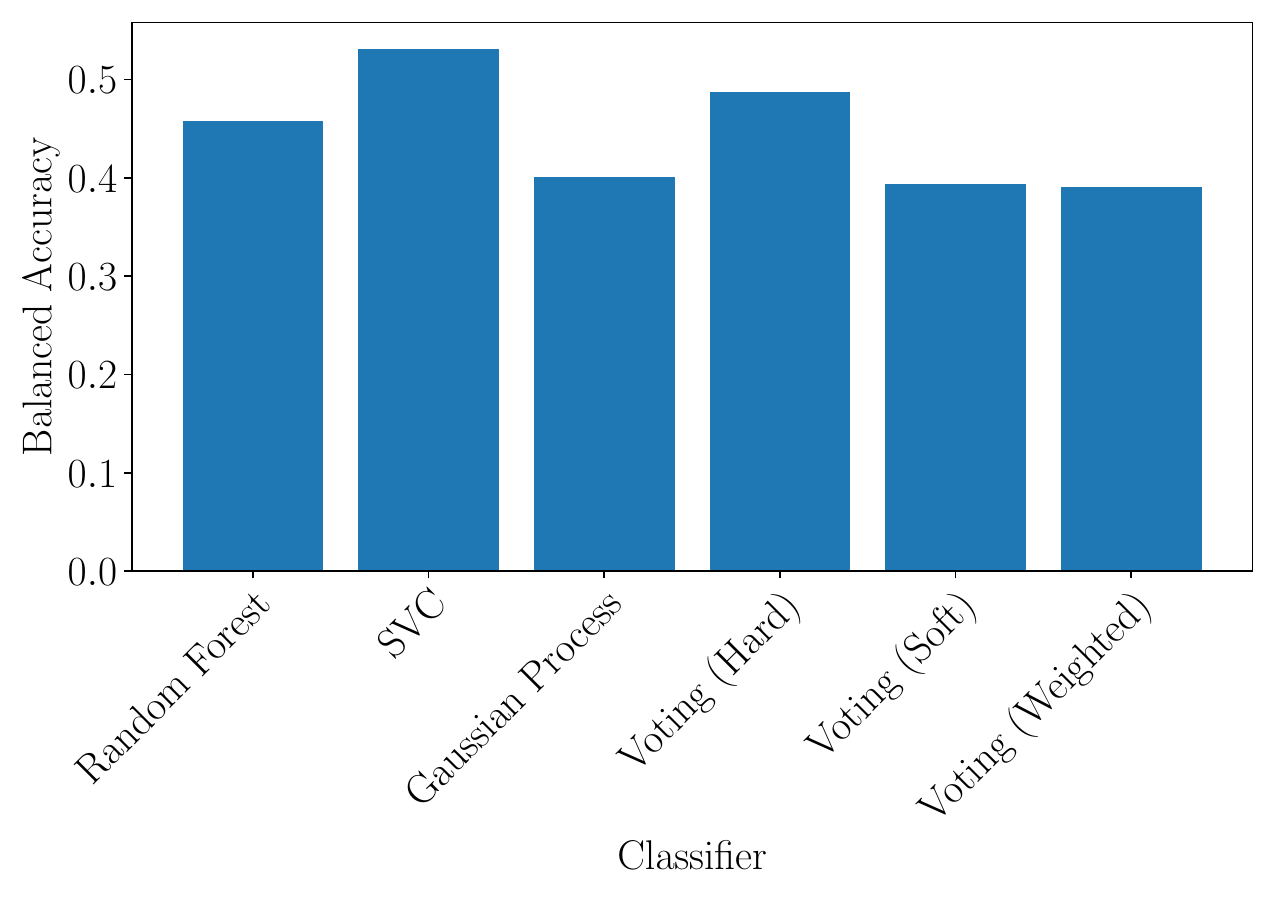}
\caption{Balanced accuracy for each optimized classifier, averaged over five foldings of the data. The SVC is the best overall, with a balanced accuracy of 0.53. Among the three voting classifiers, the Voting (Hard) classifier performs best with a balanced accuracy of 0.49, still below the SVC.}\label{fig:classifier_performance}
\end{figure}

Finally, we explore three variations of a voting classifier. Such a classifier consists of an ensemble of individual classifiers, each which ``votes'' by assigning a class to a given sample. The final assigned class can either be chosen with a ``hard'' (the class with the most votes wins) or ``soft'' (class assignments are weighted by the probabilities output by each classifier) strategy. We construct two voting classifiers that each use a different voting strategy, using RF, SVC, and GP classifiers as the individual components. We refer to these as the Voting (Hard) and Voting (Soft) classifiers. We also make a third voting classifier that also uses a soft voting strategy, but the votes from each component classifier weighted by the balanced accuracy determined via cross-validation. We refer to this as the Voting (Weighted) classifier. We score each voting classifier by averaging the balanced accuracy taken from five stratified folds of the data. Figure \ref{fig:classifier_performance} shows the balanced accuracy for the three optimized classifiers, as well as the three voting classifiers; the SVC performs ``best,'' though all classifiers return similarly low balanced accuracies between $\sim$0.4 and 0.55. Both the Voting (Soft) and Voting (Weighted) classifiers perform comparably with the worst classifier, the GP. This is due to the fact that, while the SVC often selects one individual class with high probability, both the RF and GP tend to select multiple classes with high probability (with the GP sometimes selecting all classes with roughly equal probability, usually slightly favoring the classes selected by the RF). This can result in both the RF and GP voting for the wrong class with higher probability than the correct vote from the SVC, leading to the poor observed performance.

\subsection{SVC Performance}

\begin{figure*}[ht!]
\plotone{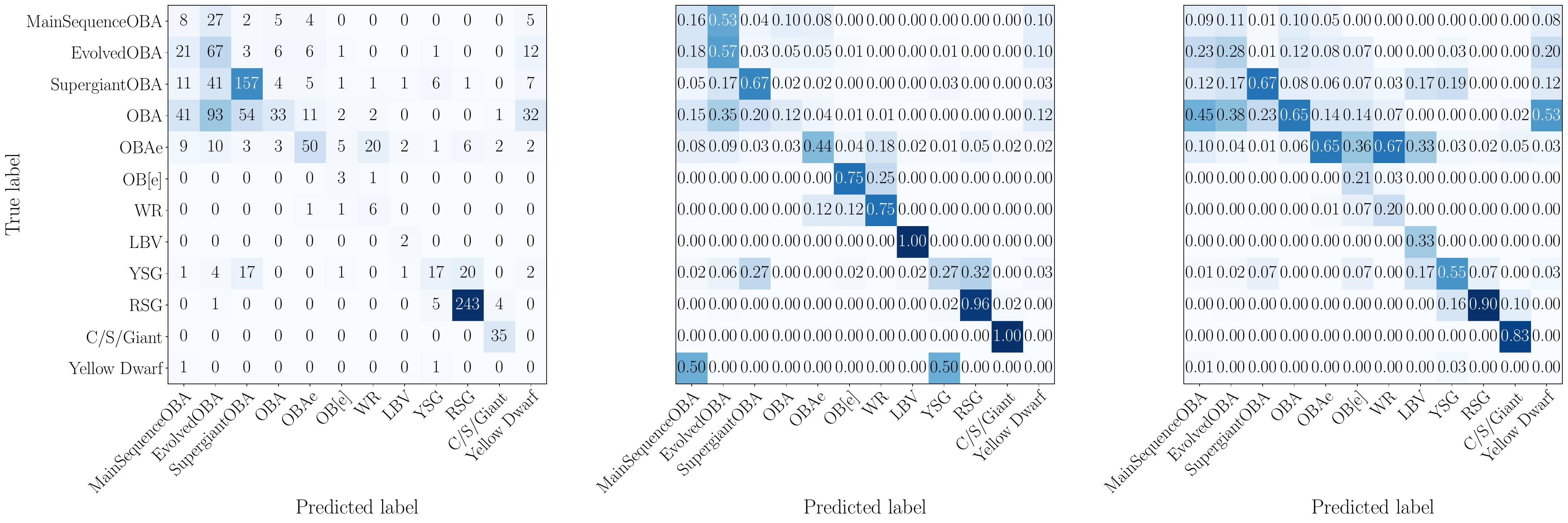}
\caption{{\it Left}: Matrix showing the number of stars in the test set with true label indicated on the y-axis that are assigned the label on the x-axis. {\it Center}: Confusion matrix for the SVC, calculated by normalizing each row of the left panel by the total number of stars in that row. Values correspond to the {\it fraction} of samples in the test set with true label indicated on the y-axis that are assigned the label on the x-axis, such that the values along the diagonal are the fraction of each class that is correctly classified. {\it Right}: Efficiency matrix for the SVC, calculated by normalizing each column of the left panel by the total number of stars in that column. Values in each box correspond to the fraction of samples in the test set assigned the label on the x-axis that belong to the class on the y-axis, such that the values along the diagonal correspond to the precision (one minus the contamination). Darker colors in all panels correspond to more/a higher fraction of stars.} \label{fig:matrix_plots}
\end{figure*}

\begin{figure}[ht!]
\plotone{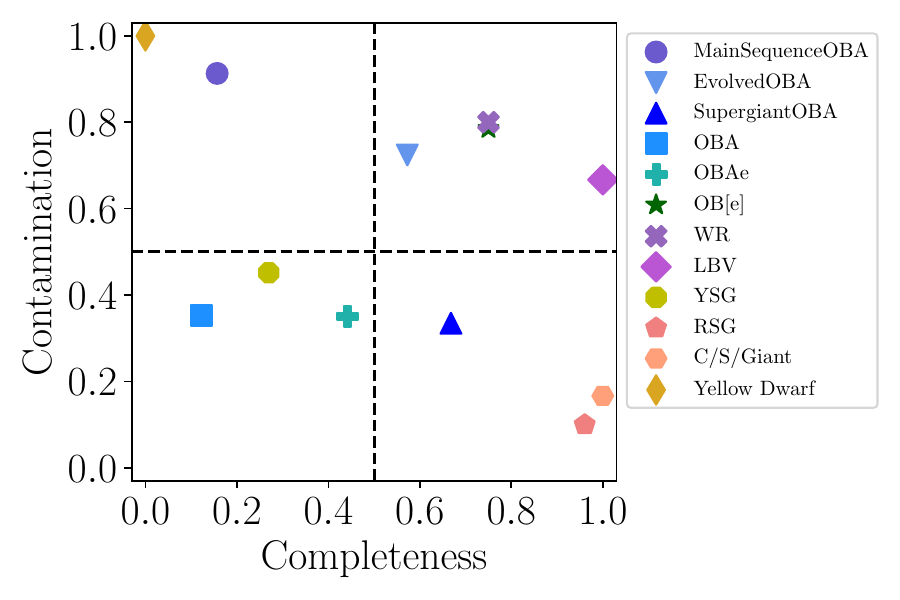}
\caption{Completeness versus contamination of each class in the test set, as classified by the SVC. A high completeness value implies members of that class are accurately classified, while a low contamination value implies that an object classified as such is likely to belong to that class. The figure is roughly divided into four quadrants; stars with classes in the bottom-right quadrant can be considered to be well-classified, in the sense that they have high completeness and low contamination.}\label{fig:completeness_contamination_SVC}
\end{figure}

 The procedure above results in values for the SVC hyperparameters of \texttt{kernel = linear}, and  \texttt{C = 0.01}. With these hyperparameters, we fit the SVC to the training set, use the \texttt{StandardScaler} and \texttt{IterativeImputer} that were previously fit to the training set to transform the test set, and use the SVC to predict the labels of the test set. The left-hand panel of Figure \ref{fig:matrix_plots} shows the raw number of stars in the test set with the true label given on the y-axis, and the predicted label given on the x-axis. The center and right-hand panels show this matrix, where each row/column is normalized by the total number of stars in that row/column, yielding the confusion/efficiency matrices, respectively. The $i,j$ entry in the confusion matrix (center panel) corresponds to the fraction of objects in the test set belonging to class $i$ (shown on the y-axis) that are assigned class $j$ (shown on the x-axis). Entries along the diagonal are the completeness (also called the recall in some contexts), i.e., the percentage of a given class that is accurately recovered by the classifier. The $i,j$ entry in the efficiency matrix (right-hand panel) is the fraction of objects in the test set classified as $j$, that belong to class $i$. Entries along the diagonal are equivalent to the precision (equivalent to one minus the contamination), i.e., the percentage of an observed class that is made up of true members of that class. Figure \ref{fig:completeness_contamination_SVC} shows the completeness versus the contamination for each class. Completeness is just the diagonal of the corresponding row/column of the confusion matrix, and contamination is one minus the diagonal of the corresponding row in the efficiency matrix.\footnote{We note that a variety of terms are used in the classification problem, some of which (i.e., completeness and contamination) are familiar to astronomy, which we briefly summarize here. The completeness (or recall) is also referred to as the true positive rate in the binary classification case. The accuracy refers to the the sum of the diagonal in the left panel of Figure \ref{fig:matrix_plots} divided by the number of objects in the test set. The contamination is also called the false positive rate in the binary classification case; the precision refers to one minus the contamination.}

The SVC performs poorly on non-supergiant OBA stars. This is perhaps unsurprising given that both the observed colors and interior structures of OBA stars as they evolve from the zero-age main sequence (ZAMS) to the terminal-age main sequence (TAMS) do not change drastically compared to the much more evolved states that we also consider. The classifier classifies OBAe stars with somewhat lower contamination compared to main sequence and evolved OBA stars, though with comparably low completeness. True OBAe stars are either misclassified as other types of OBA star or as WRs, while stars falsely labeled as OBAe are mostly true OBA stars, with the exception of one true WR. 75\% of WR stars are recovered, but only 6/30 stars identified as WRs in the test set are true WRs; given the importance of WRs for both the physics of mass loss and studying evolved massive stellar populations \citep{dornwallenstein18,dornwallenstein20}, future work will focus on developing a classifier specifically for identifying WR stars. 

All LBVs in the test set are recovered; while such high accuracy is often seen as a sign of overfitting, we choose not to focus on this subclass, given both the disputed evolutionary status of LBVs \citep{smith15,humphreys16,aadland18} and the fact that only two LBVs exist in the training set. Yellow supergiants are only classified with 27\% accuracy. As discussed in \S\ref{sec:sample} and shown in Figure \ref{fig:labelcmd}, the yellow supergiant label is assigned to stars with optical colors consistent with hot stars as well as RSGs. This is reflected in the types of stars that YSGs are mistaken for, as well as the stars that are mistaken for YSGs. Overall, the classifier performs best on the coolest stars in the sample. RSGs are classified with 96\% accuracy, and only 10\% contamination. Meanwhile the classifier performs exceptionally well at identifying low mass contaminants, at the cost of misclassifying four RSGs, two OBAe stars, and one OBA star. 

Overall, an SVC trained on these refined labels appears to have little use. With the exception of RSGs and low-mass giants, the remaining classes have low accuracy, high contamination, or both. We nonetheless use the SVC to predict labels for the 2550 stars initially labeled as ``Miscellaneous Variable'' or ``Unknown/Candidate.'' We identify 79 candidate RSGs and 36 candidate C/S/Giant stars, of which we expect $\sim$71 and 30 to be genuine, respectively, given the efficiency matrix. We list the candidate RSGs in Table \ref{tab:rsg_candidates}. A small spectroscopic observing campaign would easily confirm the ability of this classifier to correctly identify RSGs and low-mass giants.

\begin{deluxetable}{lcc}
\tabletypesize{\scriptsize}
\tablecaption{Common names and coordinates of stars predicted to be RSGs by the SVC trained on refined labels.\label{tab:rsg_candidates}}
\tablehead{\colhead{Common Name} & \colhead{R.A. [deg]} & \colhead{Dec [deg]}} 
\startdata
SP77  48-11 & 81.07900941 & $-70.43417562$ \\ 
WISE J185608.58-163255.1 & 284.03575762 & $-16.54867009$ \\ 
W61 19-14 & 83.07777354 & $-67.52941938$ \\ 
OGLE BRIGHT-LMC-MISC-169 & 72.94711333 & $-69.32348227$ \\ 
WISE J194127.64+385155.3 & 295.36520609 & $38.86536427$ \\ 
 NGC 2004   BBBC     431 & 82.69161818 & $-67.29036242$ \\ 
{[KWV2015]} J045626.51-692350.6 & 74.11062177 & $-69.39740804$ \\ 
W61  6-54 & 85.54018822 & $-69.21978048$ \\ 
WISE J064232.30-715243.3 & 100.63462144 & $-71.87871874$ \\ 
W61  6-34 & 85.51621031 & $-69.21870016$ \\ 
\enddata
\tablecomments{This table is published in its entirety in the machine-readable format. A portion is shown here for guidance regarding its form and content.}
\end{deluxetable}

\subsection{Performance on Coarse Labelling}

Examining Figures \ref{fig:matrix_plots} and \ref{fig:completeness_contamination_SVC}, we see that, while the classifier is not especially accurate except for the coolest stars, the classifier is {\it roughly} useful for sorting the test set into broad categories: different types of OBA stars are mostly (mis)classified as other classes of OBA stars; the same is true for emission line stars (OBAe, OB[e], WR, and LBV) and cool stars (YSG, RSG, C/S/Giant). 

\begin{figure}[ht!]
\plotone{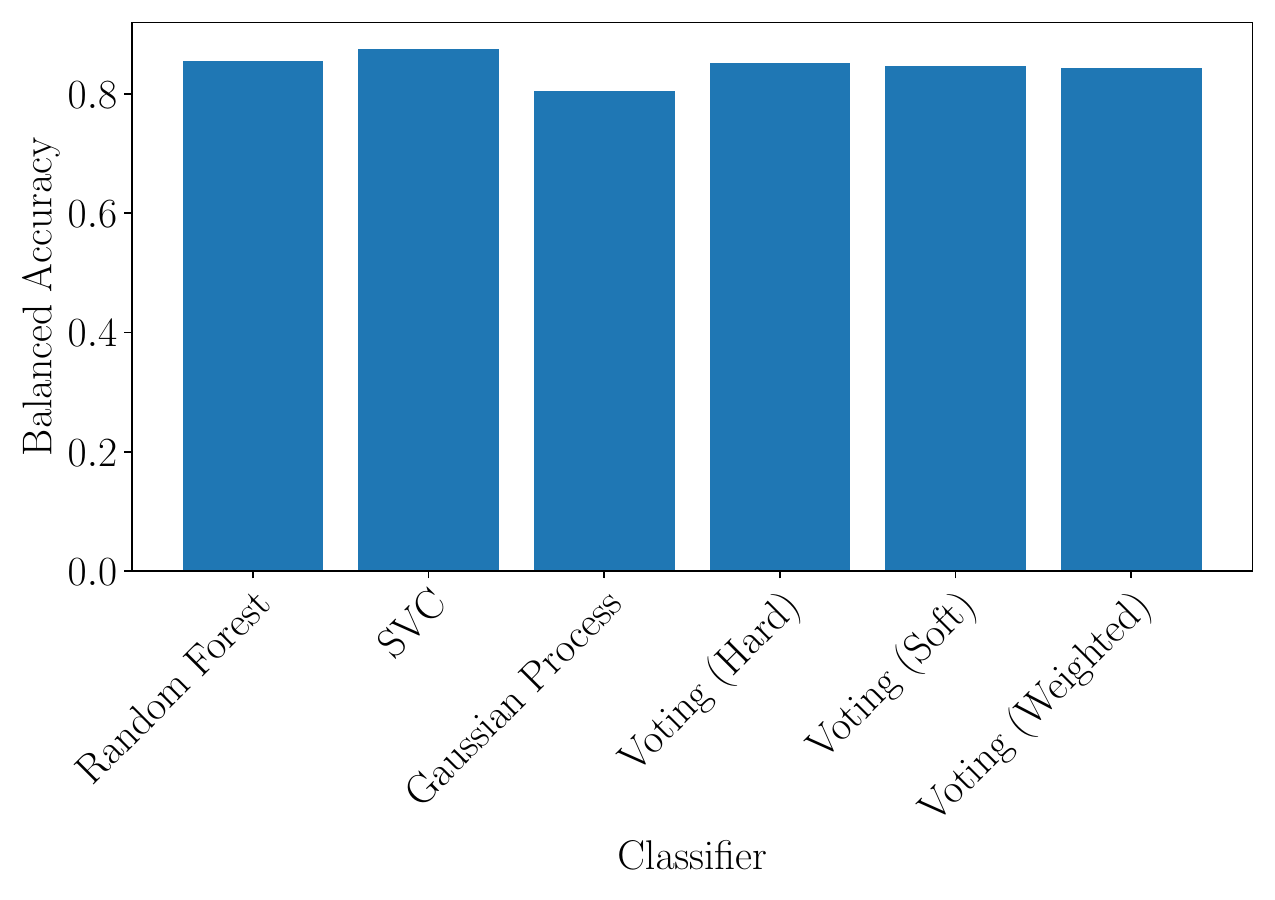}
\caption{Similar to Figure \ref{fig:classifier_performance} for classifiers trained on the coarse labels.}\label{fig:classifier_performance_coarse}
\end{figure}

For this reason, we also utilize the coarse labels introduced in \S\ref{sec:sample}. We repeat the entire process described above, beginning with the selection of the classifier. Figure \ref{fig:classifier_performance_coarse} shows the balanced accuracy for each of the classifiers discussed above, trained on the coarse labels, using a five-fold cross-validation to optimize the hyperparameters of each classifier. We find that once again, the SVC yields the highest balanced accuracy (0.876).

\begin{figure*}[ht!]
\plotone{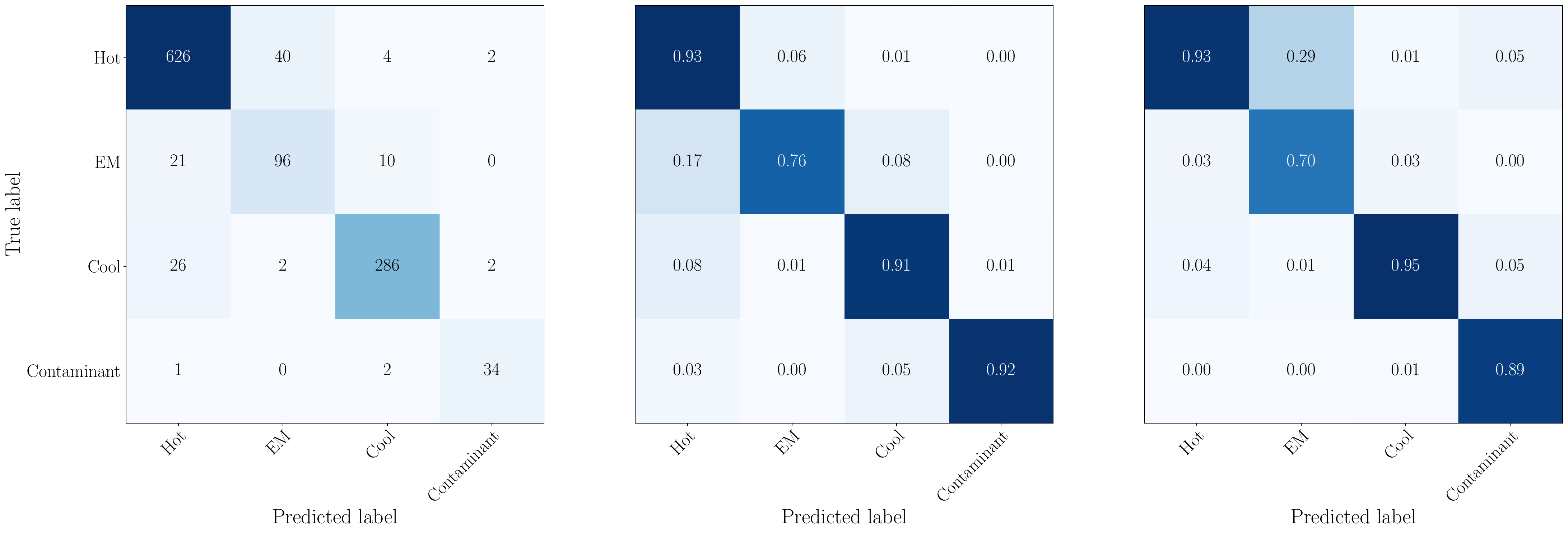}
\caption{Similar to Figure \ref{fig:matrix_plots} for the SVC trained using the coarse labels. Note that significantly more stars fall along the diagonal of each plot, reflecting the improved performance of the SVC on the coarse labels.} \label{fig:matrix_plots_coarse}
\end{figure*}

\begin{figure}[ht!]
\plotone{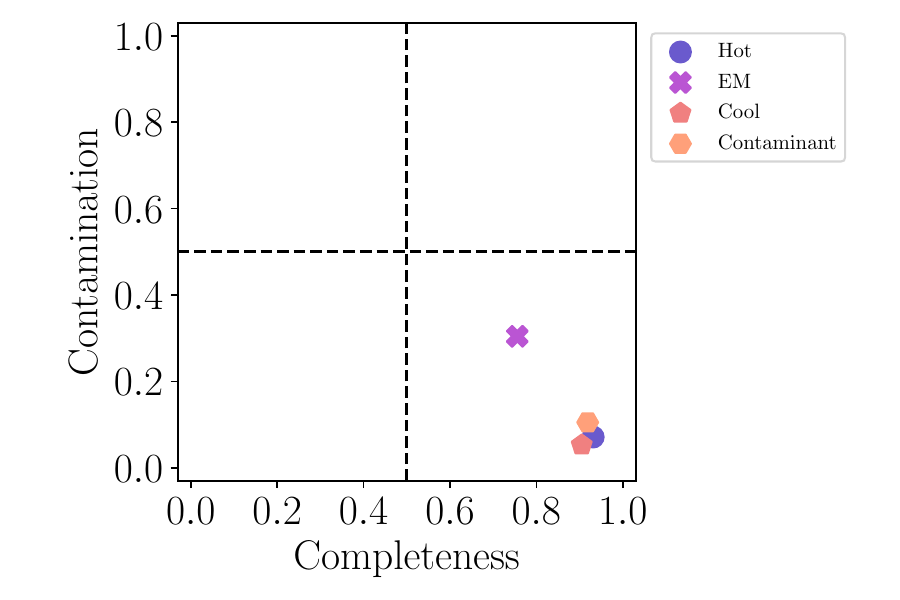}
\caption{Similar to Figure \ref{fig:completeness_contamination_SVC} for the SVC trained using the coarse labels. All coarse classes have high completeness and low contamination.}\label{fig:completeness_contamination_coarse}
\end{figure}

We keep the same scaled and imputed training and testing sets, and perform a five-fold cross-validation as before to find the optimal hyperparameters, which are \texttt{kernel=rbf},  \texttt{C = 1}, and \texttt{gamma = 1/n\_features}. With these hyperparameters, we fit the SVC to the training set before predicting labels for the test set. Figure \ref{fig:matrix_plots_coarse} shows the confusion and efficiency matrices similar to Figure \ref{fig:matrix_plots}, while Figure \ref{fig:completeness_contamination_coarse} shows the completeness versus the contamination similar to Figure \ref{fig:completeness_contamination_SVC}. All told, the SVC performs significantly better compared to the classifier trained on the refined labels, recovering all classes with $\geq75$\% completeness and $\leq30$\% contamination. 

Of the emission line stars that are correctly identified, 83 are OBAe stars, three are OB[e] stars, eight are WRs, and two are LBVs. This is 73\%, 75\%, 100\%, and 100\%, respectively of these stars that are in the test set, implying that the performance of the SVC on emission line stars is not dominated entirely by OBAe stars (which comprise the majority of emission line stars in the test set). Of the stars mislabeled as contaminants, two are OBA stars and two are RSGs. One true C/S/Giant star, and two Yellow Dwarfs are misclassified.

\begin{figure}[ht!]
\plotone{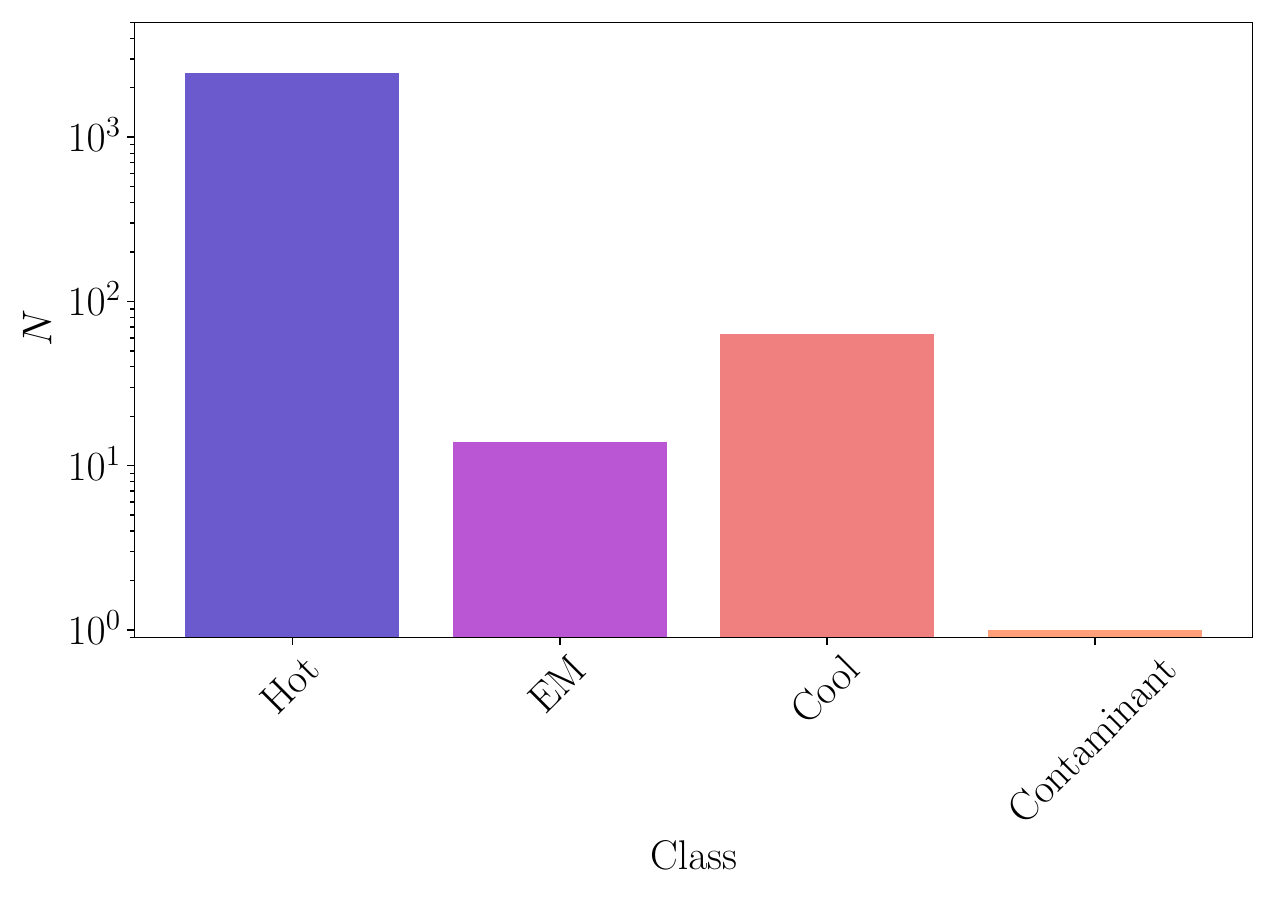}
\caption{Distribution of coarse labels assigned to 2550 stars with no previously known class.}\label{fig:unknown_labels_coarse}
\end{figure}

We then use the SVC to predict the coarse labels for the same 2550 stars as above. Figure \ref{fig:unknown_labels_coarse} shows the distribution of these predicted labels. The majority (2472 stars) are labeled as Hot. 63 stars are labeled as Cool, three of which are already identified in SIMBAD as candidate AGBs or RGBs. 14 of these stars are labeled as emission line stars, of which 9-10 are likely to actually be emission line stars, assuming 30\% contamination. We list all 2550 stars' common names, coordinates, and predicted coarse label in Table \ref{tab:coarse_candidates}.

\subsubsection{Feature Importance}\label{subsubsec:featureimportance}

We can also identify which features contribute most to the overall performance of the classifier on the coarse labels. To do this, we initialize a new \texttt{SVC} object with the same hyperparameters, and perform a ``greedy search'' over features, defined as follows: For each feature in the scaled and imputed training set, we train the SVC on just this feature across five stratified folds of the training set, and record the average and standard deviation of the balanced accuracy. We select the feature that yields the highest average balanced accuracy. This has the advantage of ensuring the contribution of each feature to the balanced accuracy is stable across subsets of the data. We then train the SVC on all combinations of this feature and the remaining features, selecting which combination again yields the highest average balanced accuracy. This process is repeated until all features are used.

\begin{figure*}[ht!]
\plotone{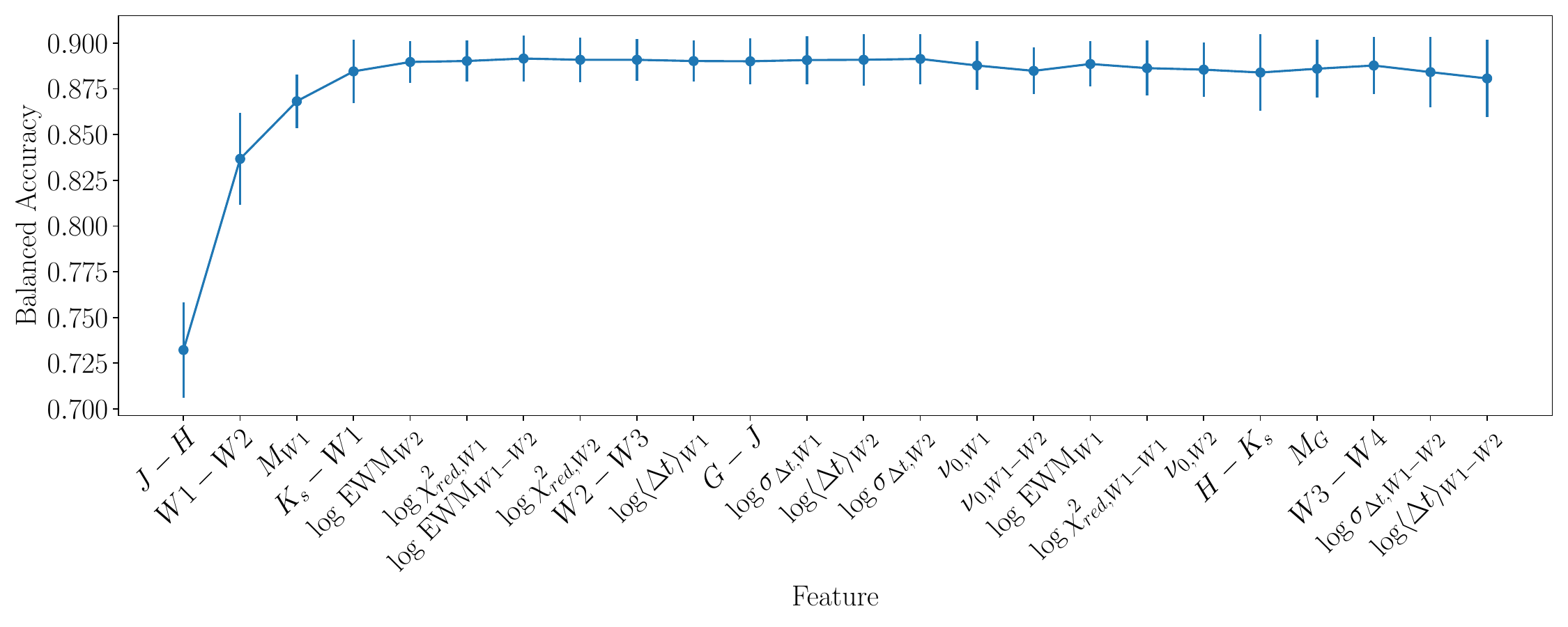}
\caption{Mean balanced accuracy of the SVC for coarse labels trained on successively added features, calculated from five stratified folds of the data. The balanced accuracy reaches a maximum after the first seven features. Errorbars indicate the standard deviation of the balanced accuracy across folds.}\label{fig:feature_importance_coarse}
\end{figure*}

Figure \ref{fig:feature_importance_coarse} illustrates this process. The x-axis shows the feature that is selected at each stage of the greedy search. The y-axis shows the mean balanced accuracy of the SVC at that stage. Errorbars show the standard deviation of the balanced accuracy across the five folds of the training set. The balanced accuracy reaches a maximum after the first seven features: $J-H$, $W1-W2$, $M_{W1}$, $K_s-W1$, $\log EWM_{W2}$, $\log\chi^2_{red,W2}$, and $\log EWM_{W1-W2}$. However, the contribution to the balanced accuracy from all but the first four features is small. This suggests that variability amplitude is a useful, though not critical metric to obtain, while variability timescales are not necessary. Finally this suggests that photometry bluer than $J$-band is also unnecessary.

We can also examine the importance of each feature for classifying individual classes. We perform the same greedy search over the features, instead calculating a performance metric that focuses on the performance on a specific class. One option is the $F_\beta$ measure:
\begin{equation}
    F_\beta = (1+\beta^2) \frac{{\rm completeness}\cdot{\rm precision}}{\beta^2\cdot{\rm completeness}+{\rm precision}}
\end{equation}
where $\beta$ is a free parameter that sets the relative importance of completeness compared to precision. Common choices are $\beta=1$ (i.e., $F_1$, a harmonic mean of completeness and precision), $\beta=0.5$, and $\beta=2$ \citep{chinchor92}. We adopt $F_2$ (i.e., $\beta=2$), because we prioritize generating complete samples of rare massive stars.

\begin{figure*}[ht!]
\includegraphics[trim=0mm 0mm 0mm 0mm, clip, width=\textwidth]{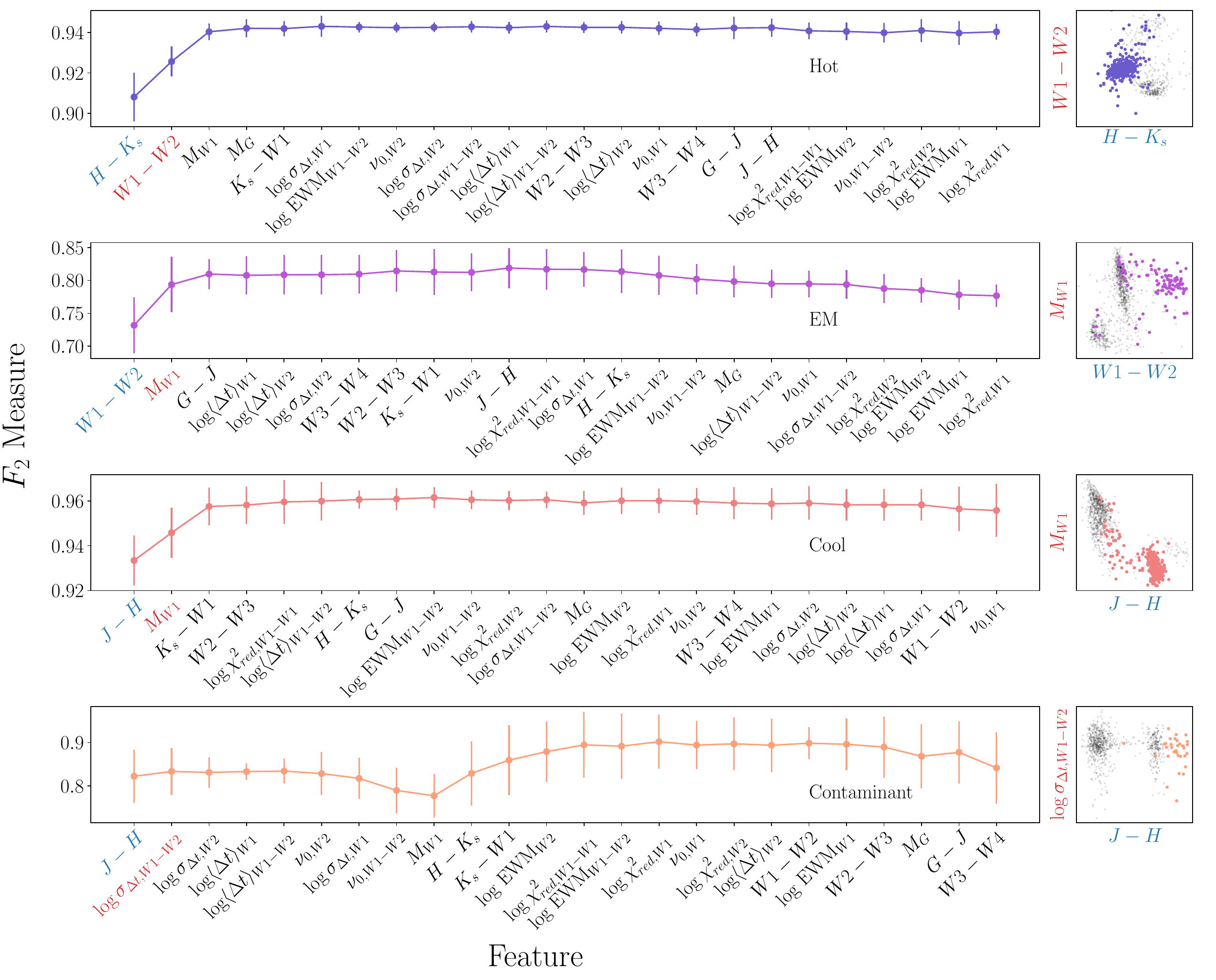}
\caption{Left panels are similar to Figure \ref{fig:feature_importance_coarse}, except using the $F_2$ measure calculated for hot stars (top), emission line stars (second panel), cool stars (third panel), and contaminants (bottom). The right panel in each row shows a scatter plot of only first and second most important features (indicated with blue and red text respectively) drawn from the test set, with stars belonging to the corresponding class in each row highlighted. Note that the features plotted are the scaled and imputed values, {\it not} the original values listed in Table \ref{tab:feature_values}.} \label{fig:feature_importance_class}
\end{figure*}

The left panels of Figure \ref{fig:feature_importance_class} show the $F_2$ measure as a function of successively added features, calculated specifically for hot stars (top), emission line stars (second panel), cool stars (third panel), and contaminants (bottom). The results for both hot and cool stars are mostly similar to the results for the overall classifier in Figure \ref{fig:feature_importance_coarse}, in the sense that the best performance is reached after including a mix of near- and mid-infrared colors and magnitudes. The main difference is that $M_G$ is the fourth most important feature for classifying hot stars.

For emission line stars, a maximum in the mean $F_2$ is reached after 11 features: $W1-W2$, $M_{W1}$, $G_J$, $\log\langle\Delta t\rangle_{W1}$, $\log\langle\Delta t\rangle_{W2}$, $\log\sigma_{\Delta t,W2}$, $W3-W4$, $W2-W3$, $K_s-W1$, $\nu_{0,W2}$, and $J-H$. However, given the errorbars, only the first three features contribute meaningfully, with the remaining features consistent with a constant value of $F_2$. Interestingly, compared to its contribution to the overall balanced accuracy of the classifier, bluer photometry (signified by the presence of $G-J$ in the above list) is much more important for identifying emission line stars. While variability metrics are included in the above list, they do not significantly contribute to the $F_2$ score.

For contaminants, a total of 15 features are required in order to maximize $F_2$: $J-H$, $\log\sigma_{\Delta t,W1-W2}$, $\log\sigma_{\Delta t,W2}$, $\log\langle \Delta t \rangle_{W1}$, $\log\langle \Delta t \rangle_{W1-W2}$, $\nu_{0,W2}$, $\log\sigma_{\Delta t,W1}$, $\nu_{0,W1-W2}$, $M_{W1}$, $H-K_s$, $K_s-W1$, $\log$ EWM$_{W2}$, $\log\chi^2_{red,W1-W1}$, $\log$ EWM$_{W1-W2}$, and $\log\chi^2_{red,W1}$. Notably, the $F_2$ measure first {\it decreases} as features are added, before increasing to the maximum after $M_{W1}$. This trend is unintuitive compared to the other panels in the Figure. It may be a result of the fact that increased features improve the precision of the classifier at the cost of completeness, resulting in a decrease in $F_2$ due to the increased weighting of completeness. 

To demonstrate the capabilities of the classifier using a limited set of features, we plot the scaled and imputed test set --- which was not used in the greedy search algorithm --- in the right panels of Figure \ref{fig:feature_importance_class}, using only the two most important features in each row. Stars belonging to the corresponding coarse class are plotted as larger, colored points, with gray points in the background corresponding to stars in test set with different coarse labels.. In all cases, most members of the test set with that label are well-separated from the other stars. As expected from the left panels, most of the separation is along the x-axis which corresponds to the most important feature, with the second-most important feature plotted on the y-axis providing some additional differentiation, especially for hot stars. In the case of contaminants, the second-most important feature provides little-to-no additional information, consistent with the lack of change in $F_2$ with increased features. 

We conclude that while small numbers of features can be used to classify hot, cool, and (remarkably) emission line stars with high accuracy and precision ($F_2\gtrsim0.9$ for hot and cool stars, and $F_2\gtrsim0.8$ for emission line stars), a large number of features is necessary in order to maximize the number of accurately identified contaminants. This includes time domain features, where the drastically different structures of old AGB and RGB stars compared to massive cool supergiants may be imprinted. Already extragalactic massive star samples are contaminated by foreground giants in the Milky Way halo; distant stars that can be resolved by {\it Webb} and {\it Roman} will have comparable brightnesses to cool dwarfs that are too faint to be filtered out using astrometry from \Gaia. Developing the infrastructure to reliably remove these contaminating objects from massive star samples will be that much more critical.

 \section{Discussion \& Conclusion}\label{sec:conclusion}

In the coming decades, space-based infrared observatories like {\it Webb} and {\it Roman} will give us access to unprecedentedly large samples of evolved massive stars. Therefore, we need to be prepared to leverage these data to search for stars in the most interesting evolutionary states. Obtaining spectroscopy of individual stars does not scale well at the size of the expected samples, while linear cuts in color-magnitude space are too simplistic and ignore emission line objects. Here we have demonstrated the promising performance of a support vector machine trained on $\sim0.5-22$ $\mu$m photometry and simple variability metrics. However, with currently available labels, we are not able to construct a classifier that performs well at the level of granularity needed for many science cases.

Our main results are summarized as follows:
\begin{itemize}
    \item We have assembled a large sample of evolved massive stars using distances from \Gaia DR2 and \citet{bailerjones18}, with high precision infrared photometry from \Gaia, 2MASS, and WISE.
    \item Using SIMBAD, we assign labels to all stars, and find that the sample contains a number of low mass contaminants.
    \item We find that, of the classification methods we applied, a support vector machine classification (SVC) algorithm is best at accurately labeling evolved massive stars. The SVC is fast, and has the added benefit that the underlying mathematics are well-understood.
    \item The SVC trained on refined labels is capable of identifying low mass red giant contaminants with high accuracy. However, the overall performance of this classifier is quite poor, and we do not recommend its use at present.
    \item The SVC trained using coarse spectral types performs better, as measured with the balanced accuracy score. We find higher completeness and lower contamination (Figures \ref{fig:matrix_plots_coarse} and \ref{fig:completeness_contamination_coarse}) compared to the SVC trained on the refined labels (Figures \ref{fig:matrix_plots} and \ref{fig:completeness_contamination_coarse}). With this classifier, we identify 14 candidate emission line stars from a sample of $\sim2500$ unlabeled stars. We plan to obtain spectroscopy of these stars to confirm our results.
    \item We find that the SVC performs equally as well with only a small subset of features. These features are mostly infrared colors and absolute magnitudes --- i.e., those least affected by reddening --- with small contributions from infrared variability metrics. However, if we change our performance metric to one that focuses on emission line stars, optimal performance of the classifier requires some red-optical photometry. We find that the added benefit of using variability metrics may not be worth the investment in telescope time in order to measure them. Of course, this is only the case for the sparsely sampled lightcurves in our sample; with the advent of the Legacy Survey  of  Space  and  Time  (LSST)  conducted  at  the  Vera  Rubin  Observatory, multi-color variability metrics can be estimated from well-sampled optical lightcurves for a significantly larger sample of evolved massive stars, and this claim can be reevaluated.
\end{itemize}

Ultimately, the performance of the SVC trained on the refined labels is poor. All stars in the sample are bright ($W1 < 14$), and the input features we use are easily measured, implying the classifier is not limited by the quality of the data. However, the labelling itself is not of sufficient accuracy, as can be seen in Figure \ref{fig:labelcmd}. Labels are derived inhomogeneously, and many are from spectroscopy that is now more than 50 years old. Unfortunately, these are the best labels available for this sample. At present, though curated lists of different subclasses of massive stars exist \citep[e.g.][]{richardson18}, no unified catalog of massive stars in our Galaxy or the Magellanic Clouds exists. 

Modern all-sky surveys have already given us access to precision photometric and spectroscopic measurements of unprecedented numbers of stars. Massive stars are bright, and so the existing data is of suitable signal to noise to perform spectroscopic classification. However, they are often excluded from analyses that provide value-added measurements like effective temperatures, surface gravities, compositions, radial velocities, and more that can be used to accurately classify massive stars. In order to prepare ourselves for the era of {\it Webb} and {\it Roman}, we must develop pipelines specifically tuned for evolved massive stars. This is especially true for the classes that are underrepresented in our dataset, i.e., rare emission line stars.

Along with better labels, more data will become available via future data releases of the \Gaia mission. The recent early third \Gaia release contains modest improvements in precision and sample size that are unlikely to affect our results given the high quality of the photometry in our sample. However, the full \Gaia DR3 will contain low-resolution spectra as well as epoch photometry for a limited number of sources, which have the potential to significantly improve the performance of a machine learning classifier. On the horizon, the Legacy  Survey  of  Space  and  Time  (LSST)  conducted  at  the  Vera  Rubin  Observatory will measure the multi-color variability of massive stars at higher cadence, while its large telescope aperture will help define a much larger sample. As we demonstrate with Figure \ref{fig:feature_importance_class}, it is possible to select features that maximize the performance of the SVC for specific classes. With a larger sample, we may be able to optimize the SVC to search for specific classes of evolved massive stars. 

\acknowledgments

The authors acknowledge that the work presented was largely conducted on the traditional land of the first people of Seattle, the Duwamish People past and present and honor with gratitude the land itself and the Duwamish Tribe.
\\
This research  was  supported  by  NSF  grant AST 1714285 awarded to EML.

JRAD and DH acknowledge support from the DiRAC Institute in the Department of Astronomy at the University of Washington. The DiRAC Institute is supported through generous gifts from the Charles and Lisa Simonyi Fund for Arts and Sciences, and the Washington Research Foundation.

DH is supported by the Women In Science Excel (WISE) programme of the Netherlands Organisation for Scientific Research (NWO).

This project was developed in part at the 2018 Gaia Sprint, hosted by the eScience and DiRAC Institutes at the University of Washington, Seattle.

This research has made use of the VizieR catalogue access tool, CDS, Strasbourg, France (DOI: 10.26093/cds/vizier). The original description of the VizieR service was published in A\&AS 143, 23. This research has made use of the SIMBAD database, operated at CDS, Strasbourg, France. This publication makes use of data products from the Two Micron All Sky Survey, which is a joint project of the University of Massachusetts and the Infrared Processing and Analysis Center/California Institute of Technology, funded by the National Aeronautics and Space Administration and the National Science Foundation.

This work made use of the following software:

\vspace{5mm}

\software{Astropy v3.2.2 \citep{astropy13,astropy18}, Astroquery v0.3.10 \citep{astroquery19}, Matplotlib v3.1.1 \citep{Hunter:2007}, makecite \citep{makecite18}, NumPy v1.17.2 \citep{numpy:2011}, Pandas v0.25.1 \citep{pandas:2010}, Python 3.7.4, Scikit-learn v0.21.3 \citep{scikit-learn11}, Scipy v1.3.1 \citep{scipy:2001}}

\appendix
\restartappendixnumbering

\section{Coarse labels for 2550 stars}\label{app:A}

Table \ref{tab:coarse_candidates} lists all 2550 stars with no known label, and predicted labels generated by the SVC trained on coarse labels.

\begin{deluxetable*}{lccr}
\tabletypesize{\scriptsize}
\tablecaption{Common names, coordinates, and predicted labels of 2550 stars input to the SVC trained on coarse labels.\label{tab:coarse_candidates}}
\tablehead{\colhead{Common Name} & \colhead{R.A. [deg]} & \colhead{Dec [deg]} & \colhead{Predicted Coarse Label}} 
\startdata
WISE J000559.28-790653.3 & 1.49713706 & $-79.11483482$ & Hot \\ 
TYC 4500-1480-1 & 2.86210879 & $79.08686958$ & Hot \\ 
BD+61    45 & 5.25504760 & $62.77064970$ & Hot \\ 
 NGC  104    LEE    2520 & 5.41170226 & $-72.21106679$ & Hot \\ 
WISE J002203.44-693554.7 & 5.51434821 & $-69.59851087$ & Hot \\ 
WISE J002207.43-742212.1 & 5.53102165 & $-74.37003199$ & Hot \\ 
WISE J002318.05-742326.4 & 5.82523611 & $-74.39068759$ & Hot \\ 
WISE J002340.20-750446.9 & 5.91756693 & $-75.07972556$ & Hot \\ 
WISE J002758.92-764527.2 & 6.99552600 & $-76.75757402$ & Hot \\ 
WISE J002759.32-742119.8 & 6.99734043 & $-74.35552728$ & Hot \\ 
\enddata
\tablecomments{This table is published in its entirety in the machine-readable format. A portion is shown here for guidance regarding its form and content.}
\end{deluxetable*}

\newpage
\bibliography{bib}
\bibliographystyle{aasjournal}

\end{document}